\documentclass[journal]{IEEEtran}

\usepackage{color}
\usepackage{graphicx}
\usepackage{subcaption}
\usepackage{amsmath}
\usepackage{amssymb}
\usepackage{booktabs}
\newcommand{\red}[1]{\textcolor{black}{#1}}

\newcommand{\vect}[1]{\boldsymbol{#1}}

\newcommand\revcol{\color{black}}

\usepackage{tikz,pgfplots}
\usepgfplotslibrary{groupplots}
\pgfplotsset{compat=1.14}
\usetikzlibrary{spy}
\newcommand\x{0.22}
\newcommand\y{1.5cm}
\newcommand\locroiX{-1}
\newcommand\locroiY{0.5}
\newcommand\locmagX{1.95}
\newcommand\locmagY{1.2}

\usepackage{stfloats}

\usepackage{url}

\begin{document}

\title{Multi-modal Deep Guided Filtering for Comprehensible Medical Image Processing}

\author{
	{Bernhard~Stimpel, Christopher~Syben, Franziska~Schirrmacher, Philip~Hoelter, Arnd~D\"orfler,~and~Andreas~Maier,~\textit{Member,~IEEE}}%
\thanks{B.~Stimpel, C.~Syben, F.~Schirrmacher, and~A.~Maier are with Pattern Recognition Lab, Friedrich-Alexander~University Erlangen-Nuremberg, Erlangen, Germany. (email: bernhard.stimpel@fau.de)}%
\thanks{P.~Hoelter and A.~D\"orfler are with University Hospital Erlangen, Erlangen, Germany.}%
\thanks{Financial support for this project was granted by the Emerging Fields Initiative (EFI) of the Friedrich-Alexander~University Erlangen-Nuremberg as well as Siemens Healthineers. Furthermore, we thank the NVIDIA Corporation for their hardware donation. }%
}
\makeatletter
\def\blfootnote{\gdef\@thefnmark{}\@footnotetext}
\makeatother

\maketitle
\begin{abstract}
Deep learning-based image processing is capable of creating highly appealing results. However, it is still widely considered as a "blackbox" transformation. In medical imaging, this lack of comprehensibility of the results is a sensitive issue. The integration of known operators into the deep learning environment has proven to be advantageous for the comprehensibility and reliability of the computations. Consequently, we propose the use of the locally linear guided filter in combination with a learned guidance map for general purpose medical image processing. 
The output images are only processed by the guided filter while the guidance map can be trained to be task-optimal in an end-to-end fashion. We investigate the performance based on two popular tasks: image super resolution and denoising. The evaluation is conducted based on pairs of multi-modal magnetic resonance imaging and cross-modal computed tomography and magnetic resonance imaging datasets. 
For both tasks, the proposed approach is on par with state-of-the-art approaches. Additionally, we can show that the input image's content is almost unchanged after the processing which is not the case for conventional deep learning approaches. On top, the proposed pipeline offers increased robustness against degraded input as well as adversarial attacks. 
\end{abstract}

\section{Introduction}
\label{sec:Introduction}

\blfootnote{Copyright (c) 2019 IEEE. Personal use of this material is permitted. However, permission to use this material for any other purposes must be obtained from the IEEE by sending a request to pubs-permissions@ieee.org.}
Machine learning has experienced a rapid growth in medical image processing in recent years. Especially with the advent of deep learning, the possibilities in many applications have been substantially expanded \cite{Krizhevsky2012,LeCun2015, Gatys2016,Ronneberger2015}.
%comprehensibility
However, deep learning methods imply a high-dimensional, non-linear transformation of the input data within the networks, which is difficult to comprehend from the outside \cite{Koh2017,Shwartz-Ziv2017, Carlini2017,Antun2019}. While this can be tolerated in many areas of research due to the empirically good results, there is only a small margin of tolerance for possible errors in the processing of medical image data. 
This is particularly fatal in applications in which the image information is fundamentally altered such as image reconstruction, super resolution, or denoising. It is therefore difficult to differentiate between previously existing and subsequently added or removed information.
%known operators
Consequently, a number of recent studies have focused on the combination of learning-based methods and known operators \cite{Wurfl2016,Adler2018,Hammernik2018,Maier2019}. The latter are clearly defined by known mathematical formulations and can provide the necessary level of comprehensibility while still taking advantage of the many benefits of data-driven learning. 
Following these efforts, we present the well-known guided image filter \cite{He}, a locally linear operating filter, paired with an end-to-end learned guidance map \cite{Wu2018}. The guided filter was successfully applied to a variety of tasks in the past \cite{He,ShutaoLi2013,Kou2015}. Usually, the guidance map in these applications is represented by an existing image, similar to the one to be processed, that exhibits the desired characteristics. Leveraging the ability of neural networks to extract the most valuable information from data  \cite{Wu2018}, the corresponding guidance map can be trained in a task-optimal fashion instead.
Thereby, an exploitable aspect in medical imaging is that often multiple datasets of the same patient are available, e.g., in the form of multiple MR sequences, PET-CT scans, or novel hybrid CT-MR imaging modalities \cite{Wang, Fahrig2001}.
Thus, using redundant information across the datasets can help to restore lower-quality or -resolution scans even if it is acquired by a different modality. 
This aids to alleviate the intrinsic problem of single image-based image restoration which is the hallucination of information that is not present in the original image. 
Combined with the guided image filter, that is employed to decouple the network's output from the image to be processed, we aim to provide a comprehensible approach to medical image enhancement.
\red{Previous research took a similar methodological direction~\cite{Stimpel2019}. However, this work neglects a thorough evaluation and analysis of the results with regard to the proclaimed comprehensibility. Furthermore, no comparison with state-of-the-art networks and the influence of the guided filter on these methodological advancements, and no ablation study is conducted.}

In the following we show that: (1) The combination of the guided filter with an end-to-end learned guidance map from multi-modal input is capable of producing results that are on par with start-of-the-art approaches on multiple tasks. (2) This can be achieved with considerably less manipulation of the underlying input image's content compared to the approaches without the guided filter. (3) The proposed pipeline is also more robust against degradations of the input as well as adversarial attacks.

We demonstrate this based on two popular tasks in natural and medical imaging processing: image super resolution (SR) \cite{Oktay2016,Pham2017,Yu2018} and denoising \cite{Gondara2016,Amiot2016,Zhang2017}. 

\section{Methods}
\label{sec:Methods}
\begin{figure*}[t]
	\centering
	\includegraphics[width=1\textwidth]{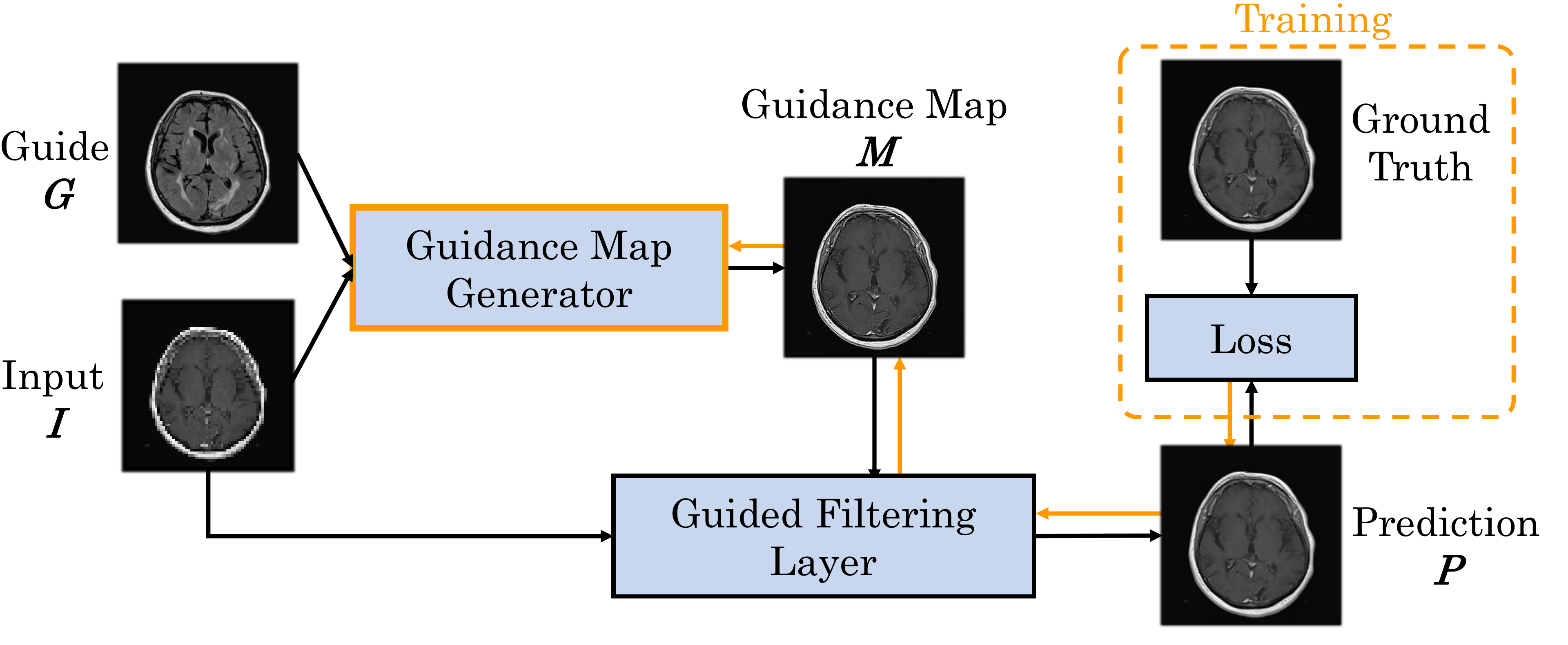}	
	\caption{The proposed guided filtering pipeline during training. Black arrows indicate the order of processing steps and orange arrows the gradient flow.
		$\vect{I}$ is the image to be processed and $\vect{G}$ the second modality guide image. The guidance map generator is represented by a neural network and transforms these to create the guidance map $\vect{M}$. 
		The input $\vect{I}$ and the guidance map  $\vect{M}$ are, subsequently, processed by the guided filter layer yielding the output  $\vect{P}$. The guided filter is used at this point to decouple the network from the image to upsample. While training of the guidance map generator, the last step is to compute the difference between the predicted output $\vect{P}$ and it's ground truth and to optimize the network using the resulting gradient. }
	\label{fig:gf_pipeline}
\end{figure*}

The proposed pipeline is formed by an image generator network, a guided filtering layer \cite{He,Wu2018} and an objective function (see Fig. \ref{fig:gf_pipeline}). 
As initially motivated, we expect multi-modal input data in the form of an image to be processed $\vect{I}$ and a complementary guide image $\vect{G}$ of arbitrary modality. These inputs are processed by the guidance map generator network to output a fused representation, the needed guidance map $\vect{M}$. The input  $\vect{I}$ and the guidance map  $\vect{M}$ are subsequently processed by the guided filter layer, yielding the predicted output $\vect{P}$. The guided filter is used at this point to decouple the network from the image to process, since we have no means to comprehend the high-dimensional, non-linear transformations of the generator network. While training of the guidance map generator, the last step is to compute the difference between the predicted output $\vect{P}$ and it's ground truth, and to optimize the network using the resulting gradient.

\subsection{Guided Image Filtering}
\label{sec:methods_gf}
The main assumption of the guided image filter is a locally linear relationship between an input image $\vect{I}$ and a guidance map $\vect{M}$ \cite{He}. It is an edge-preserving smoothing filter, i.e., information in the input image is conserved. Image information of the guidance map is transferred to the input image if and only if the underlying local information is correlating. Correlation is defined by image statistics that are determined based on the inputs to the guided filter. For a given input $\vect{I}$ and guidance map $\vect{M}$, these statistics are computed in a window $\vect{w}_k$ centered around the pixel $k$ by
\begin{equation}
\label{eq:guided_filter_ak}
a_k = \frac{\frac{1}{|\vect{w}|} \sum_{i\in \vect{w}_k} M_i I_i - \bar{M}_k \bar{I}_k}{{\sigma_{M_k}^2} + \epsilon}
\end{equation}
\begin{equation}
\label{eq:guided_filter_bk}
b_k = \bar{I}_k - a_k \bar{M}_k  \;,
\end{equation}
with $\bar{M}_k$ and $\bar{I}_k$ being the mean of $\vect{M}$ and $\vect{I}$ in the window $\vect{w}_k$, respectively. The variance of $\vect{M}$ in the currently considered window $\vect{w}_k$ is denoted by  ${\sigma_{M_k}^2}$. The size of the kernel is defined by the chosen radius $r$ which is subject to parameter tuning. The linear coefficients $a_k$ and $b_k$ are subsequently used to compute the output of the guided filter for a window $\vect{w}_k$ by
\begin{equation}
P_i = a_k M_i + b_k \;,\; \forall i \in \vect{w}_k\;.
\label{eq:guided_filter}
\end{equation}
As each image point is covered by multiple kernel windows $ \vect{w}_k$, the outputs for each point are ultimately averaged. Technically, the whole procedure can be implemented with box filter in a desirable runtime of O(N). For a more in-depth mathematical formulation and associated interpretations we kindly refer to \cite{He}. 

There is no prior on the nature of $\vect{M}$ that is expected explicitly. Though, in order to fully leverage the power of the guided filter a suitable guidance map is required. In the underlying case of multi-modal input data, a large amount of complementary information is available. This points towards a combination of the single modalities into one guidance map. As the optimal combination w.r.t. the operations performed by the guided filter is unknown, we propose to learn this directly from the available data. Fused images are unavailable by default, which renders direct supervised learning impossible.
\red{Wu et al. showed that the guided image filter can be integrated in existing deep learning environments in the form of a differentiable layer~\cite{Wu2018}. For details about the backward procedure, we kindly refer the interested reader to Algorithm~1 and Figure~2 in~\cite{Wu2018}. }
This opens up the possibility to compute a task-based loss, i.e., in the underlying case the deviation between the predicted and the ground truth image. The resulting gradient can then be propagate through the guided filtering layer to deeper levels, in this case a generator for the guidance map. The guidance map $\vect{M}$ is then defined by the mapping $\vect{M} = \phi(\vect{I}, \vect{G})$ where $\phi$ is the transform applied by the guidance map generator. 

\subsection{Guidance Map Generator: } 
\label{sec:methods_guidance_generator}
The key component in this setup is an image generator that takes multi-modal input images and outputs a task-optimal fused representation which is the needed guidance map. 
Convolutional neural networks (CNN) have shown great success in extracting the most valuable information directly from data. Consequently, a CNN is employed to generate the guidance map. 

We target two popular tasks, image super resolution and denoising. For the sake of reproducibility, proven network architectures are used to generate the guidance maps for both. The only modifications result from the need of processing multiple input images. Naturally, the addition of more guide images would be possible and is only limited by availability and computational resources. To investigate the influence of the chosen network architecture for the guided filtering process, we employ two different networks for the task of super resolution.

\subsubsection{Denoising}
\label{sec:methods_generator_denoising}
We adopt the frequently used U-net architecture \cite{Ronneberger2015} as our guidance map generator for image denoising. This network has been proven suitable for many tasks in medical image processing. We modify the original architecture to use two separate encoding paths, one for each input modality, that are fused by concatenation after the last encoding layer. Additionally, the originally used deconvolution layers are replaced by bilinear upsampling and subsequent convolution. For a graphical representation of our modifications please refer to Appendix A.

\subsubsection{Super Resolution (SR)}
\label{sec:methods_generator_sr}
In addition to the aforementioned U-net, the WDSR network \cite{Yu2018,Fan2018} is used for the super resolution task. The network is sophisticated for super resolution purposes and ranked 1\textsuperscript{st} in one of 2018's NTIRE super resolution challenges. We use the network as provided by the authors with only a few modifications to adapt it to our requirements. To accept two inputs modalities, we add a second residual connection for the high-resolution input modality. In addition, the original network was preceded by twyo convolutional encoding blocks as used in the U-net to bring the high-resolution input modality down to the same resolution as the image to be upsampled. All RGB color-related shift and scaling operations were removed. Details of the WDSR network and our minor modifications are presented in Appendix B. By including a second, task-specific network, we aim to examine the influence of the choice of the network architecture for the proposed pipeline. 

Deviating from the pipeline shown in Fig. \ref{fig:gf_pipeline}, the input for the guided filter in the super resolution case is not the low-resolution input image, but a bilinear upsampled version of it, further denoted as $\vect{I}_{\text{up}}$. This serves as an initialization for the upsampling process.

\section{Experiments}
\label{sec:exp}
The networks are trained using a  VGG-19-based \cite{Simonyan2015} feature matching (FM) loss \cite{Johnson2016} and a batch size of one. Optimization is performed using the ADAM optimizer \cite{Kingma2014} with an initial learning rate of $1\text{e}^{-5}$ and learning rate decay down to $1\text{e}^{-6}$ if no further improvement in the validation loss is observable. All network parameters are chosen such that the full GPU memory is used to ensure similar capacity. 
\begin{table*}[b!]
	\centering
	\caption{SR: quantitative results for both datasets. Bold font denotes the best result.}
	\label{tab:eval_patients}	
	\begin{tabular}{@{\extracolsep{0.1cm}}lllllll@{}}
		\toprule
		\multicolumn{7}{l}{\bfseries X-ray \& MRI Projection Images \bfseries ($128\,\text{x}\,128 \rightarrow 512\,\text{x}\,512$)}   \\ \midrule
		& Bicubic & \revcol Guided Filter &U-net w/ GF & U-net w/o GF & WDSR w/ GF & WDSR w/o GF \\
		MAE  & $0.0584\red{}\scriptstyle\pm\red{}0.1082$  & \revcol $0.0027\scriptstyle\pm0.0057$ &$0.0017\red{}\scriptstyle\pm\red{}0.0039$  & $0.0017\red{}\scriptstyle\pm\red{}0.0038$  & \boldmath$0.0017\red{}\scriptstyle\pm\red{}0.0037$  & $0.0017\red{}\scriptstyle\pm\red{}0.0051$    \\
		SSIM  & $0.8807\red{}\scriptstyle\pm\red{}0.0093$   & \revcol$0.9826\scriptstyle\pm0.0014$ &$0.9906\red{}\scriptstyle\pm\red{}0.0014$  & $0.9909\red{}\scriptstyle\pm\red{}0.0012$  & $0.9910\red{}\scriptstyle\pm\red{}0.0011$  & \boldmath$0.9912\red{}\scriptstyle\pm\red{}0.0013$  \\
		%	&         &              &             &             &            \\  
		\midrule
		\multicolumn{7}{l}{\bfseries Tomographic T1 \& T2 MRI images \bfseries ($64\red{}\text{x}\red{}64 \rightarrow 256\red{}\text{x}\red{}256$)}         \\ \midrule
		& Bicubic & \revcol Guided Filter &U-net w/ GF & U-net w/o GF & WDSR w/ GF & WDSR w/o GF \\
		MAE & $0.0525\red{}\scriptstyle\pm\red{}0.0994$  & \revcol $0.0179\scriptstyle\pm0.0414$ & $0.0098\red{}\scriptstyle\pm\red{}0.0205$  & $0.0082\red{}\scriptstyle\pm\red{}0.0184$  & $0.0076\red{}\scriptstyle\pm\red{}0.0161$  &\boldmath $0.0021\red{}\scriptstyle\pm\red{}0.0039$  \\
		SSIM & $0.7125\red{}\scriptstyle\pm\red{}0.0698$ & \revcol  $0.8697\red{}\scriptstyle\pm\red{}0.0480$ & $0.9552\red{}\scriptstyle\pm\red{}0.0173$  & $0.9740\red{}\scriptstyle\pm\red{}0.0107$  & $0.9658\red{}\scriptstyle\pm\red{}0.0099$  & \boldmath$0.9963\red{}\scriptstyle\pm\red{}0.0010$      \\ \bottomrule
	\end{tabular}
\end{table*}

\begin{table*}[b!]
	\centering
	\caption{Denoising: quantitative results for the X-ray \& MRI Projection images.}
	\label{tab:eval_patients_denoising}	
	\begin{tabular}{@{\extracolsep{0.0cm}}lcccccc@{}}
		\toprule
		%\multicolumn{6}{l}{\bfseries Denoising: X-ray \& MRI Projection Image  }   \\ \midrule
		 & \multicolumn{2}{c}{Low noise} & \multicolumn{2}{c}{Medium noise}   & \multicolumn{2}{c}{Strong noise}   \\
		 & MAE [\%] & SSIM & MAE [\%] & SSIM & MAE [\%] & SSIM   \\ \midrule
		\revcol
		Guided Filter  &
		 \revcol $0.219\scriptstyle\pm0.517$  & \revcol $0.987\scriptstyle\pm0.003$  &
		 \revcol $0.256\scriptstyle\pm0.560$  & \revcol $0.982\scriptstyle\pm0.003$  &
		 \revcol  $0.304\scriptstyle\pm0.629$ & \revcol $0.977\scriptstyle\pm0.002$   \\
		U-net w/ GF  &
		 $0.170\scriptstyle\pm0.330$  & $0.991\scriptstyle\pm0.001$  &
		  $0.231\scriptstyle\pm0.456$  & $0.986\scriptstyle\pm0.002$  &
		   $0.286\scriptstyle\pm0.564$ & $0.979\scriptstyle\pm0.002$   \\
		U-net w/o GF  &
		 $0.177\scriptstyle\pm0.336$  & $0.992\scriptstyle\pm0.001$  &
		   $0.273\scriptstyle\pm0.508$  & $0.986\scriptstyle\pm0.002$  & 
		    $0.289\scriptstyle\pm0.554$ & $0.983\scriptstyle\pm0.002$   \\
		\bottomrule
	\end{tabular}
\end{table*}

\subsection{Data}
\label{sec:exp_data}
We analyze the performance of the proposed approach based on 21 pairs of clinical patient head datasets. 
First, 8 pairs of tomographic T1 and T2 Flair MRI datasets ($256\;\text{x}\;256$) are used. Second, cone-beam X-ray and MRI projection images ($512\;\text{x}\;512$) generated from 13 patients are utilized (MR: 1.5\,T MAGNETOM Aera / CT: SOMATON Definition, Siemens Healthineers, Erlangen / Forchheim, Germany).
Two pairs of patient dataset of each modality were reserved for validation and testing. Image registration of the corresponding datasets is performed using 3D~Slicer~\cite{Pieper}. The forward projections are based on the work of hybrid MR/X-ray imaging by \cite{Stimpel2017a,Syben2018,Lommen2018} and are created with the CONRAD framework~\cite{Maier2013}. 108 projections are created per patient that are distributed equiangularly along the azimuthal and in 60$^\circ$ degree range along the inclination angle. 
Note that the type of data presented here is only to be seen as an example. The proposed approach is not tailored to any particular data type or combination of data types.

For the task of super resolution, low-resolution images are created by nearest neighbor downsampling by a factor of 4. For denoising, noisy images are generated from the ground truth by applying Poisson noise. Both methods are only an approximation of the real physical processes. In MRI, for example, the resolution and the signal-to-noise ratio are directly related \cite{Redpath1998}. Also, noise in X-ray imaging is not just Poisson noise but a combination of multiple sources of noise \cite{Hariharan2018}. For detailed investigations on a specific problem area a corresponding simulation should therefore be conducted.
\noindent
\subsection{Evaluation}
\label{sec:exp_eval}
Both tasks are evaluated quantitatively by computing the mean absolute error (MAE) and structural similarity (SSIM) measures. To provide an ablation study, all networks are evaluated with and without the guided filter, which is in the following abbreviated as \textit{w/ GF} and \textit{w/o GF} , respectively. In the latter case the networks directly yield the desired output, i.e., the guided filtering step in Fig. \ref{fig:gf_pipeline} is skipped. \red{Please note that all networks were retrained from scratch without the guided filter in order to do so. In addition, the performance of the guided filter without the end-to-end learned guidance map is evaluated, which will be denoted as \textit{only GF}. In this case the guide image $\vect{G}$ is used as guidance map $\vect{M}$  for the guided filter.}
Pixels outside the head are ignored in the course of the evaluation to avoid optimistic bias due to the large background areas. To allow for qualitative evaluation and an assessment of the visual appearance, output images for each task are presented. 
Furthermore, the influence of the guided filter's parameters, the radius $r$ of the kernel window $\vect{w}$ and the corresponding $\epsilon$ (see Eq. \ref{eq:guided_filter_ak}-\ref{eq:guided_filter}) will be investigated.

\subsection{Comprehensibility}
\label{sec:exp_comprehensibility}
Comprehensibility of the generated output is an important topic, especially in medical applications. For most proposed approaches, this is currently not possible or only possible to a limited extent \cite{Koh2017,Shwartz-Ziv2017, Carlini2017,Antun2019}. We therefore conduct three experiments regarding the comprehensibility of the results generated by the proposed guided filtering pipeline in comparison with the traditional approach solely based on a CNN. In detail, these target the preservation of the input image's content, the robustness of the pipeline against distorted input signals, and the effect of adversarial attacks on the generated results.  

\subsubsection{Content Preservation}
\label{sec:exp_content_pres}
Important image information such as regular anatomical structures but also lesions and similar pathologies is stored in the low-frequency band. 
Deviation of the processed outputs from the  input images in the low-frequency components is critical in such a sensitive field as medical imaging and any manipulation during processing must be considered with caution. A change of the high frequency bands is, however, inevitable for the application case of super resolution and denoising.  
Diagnoses that are dependent on this kind of information should be approached from the outset with a corresponding higher quality acquisition. 
Consequently, we analyze the change in low frequencies between the super-resolved and the input images or the denoised with the label images, respectively. 
To this end, the images are decomposed into different frequency parts by two-times wavelet decomposition (Daubechies-4) and all high-frequency information is discarded.
The resulting decomposition of the processed images should be as close as possible to their counterpart, as we aim to avoid a change in these frequency bands. An example of this wavelet decomposition is given in Appendix~C.

\subsubsection{Insusceptibility to Disturbances}
\label{sec:exp_disturbances}
As a second experiment, we simulate the behavior of the pipeline in the face of disturbed input data. Information that is not existing in the low-resolution images should be extracted from the supplementary guide image. If the information in the guide is distorted or not complementary to the input image the outcome should, ideally, not be influenced by this. To investigate the robustness, we manipulate the high-resolution guide images at the time of the inference with successively increasing additive zero-mean Gaussian noise \cite{Antun2019} and observe the effect on the predicted output images. 

\subsubsection{Robustness against Adversarial Attacks}
\label{sec:exp_adversarial}
Vulnerabilities of neural networks to adversarial attacks have recently been shown for many tasks \cite{Antun2019,Yuan2017}. In a third experiment, we therefore investigate the possibility of an adversarial attack on the network both without and with the guided filter. A simple attack is chosen to this end. We seek to find an additive residual to both input modalities that causes a large deviation of the prediction from the ground truth and simultaneously exhibits a small norm itself. For a detailed description we kindly refer to Appendix D. Ideally, the guided filter should limit the influence of the adversarial attack on the guidance map by decoupling it from the processed output. 

\section{Results}
\label{sec:results}

\begin{figure*}
	\input{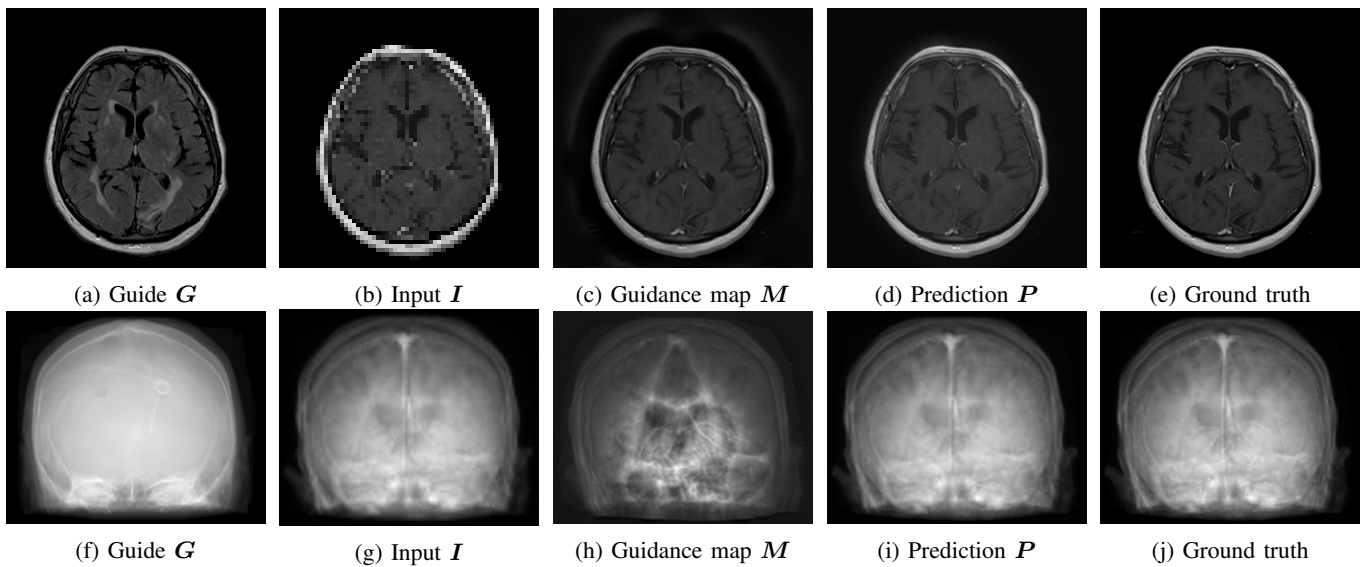}
	\caption{Inputs and outputs of the guided filtering pipeline based on the WDSR network. T1 \& T2 MRI pairs (a)-(e) and CT \& MRI projection images (f)-(j).}
	\label{fig:pipeline_io_sr}
\end{figure*}

\begin{figure*}
	\hfil
\renewcommand\x{1}
\renewcommand\locroiX{-0.7}
\renewcommand\locroiY{0.4}
\renewcommand\locmagX{1.7}
\renewcommand\locmagY{0.96}

\begin{subfigure}{0.19\textwidth}
	\begin{tikzpicture}[spy using outlines={rectangle,orange,magnification=2.5, 
		height=\y, width=\y, connect spies, /.append style={thick}}, every node/.append style={inner sep=0}]
		\node {
			\pgfimage[width=\x\textwidth]{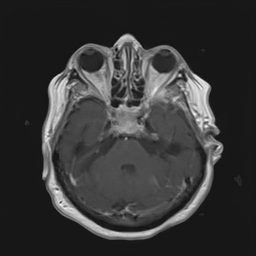}
		};%
		\spy on (\locroiX, \locroiY) in node [left] at (\locmagX, \locmagY);     
	\end{tikzpicture}
	\caption{U-net w/o GF}
\end{subfigure}	%
\hfil
\begin{subfigure}{0.19\textwidth}
	\begin{tikzpicture}[spy using outlines={rectangle,orange,magnification=2.5, 
		height=\y, width=\y, connect spies, /.append style={thick}}, every node/.append style={inner sep=0}]
		\node {
			\pgfimage[width=\x\textwidth]{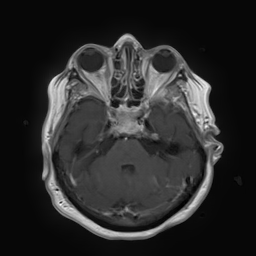}
		};
		\spy on (\locroiX, \locroiY) in node [left] at (\locmagX, \locmagY);     
	\end{tikzpicture}
	\caption{U-net w/ GF}
\end{subfigure}	%
\hfil
\begin{subfigure}{0.19\textwidth}
	\begin{tikzpicture}[spy using outlines={rectangle,orange,magnification=2.5, 
	height=\y, width=\y, connect spies, /.append style={thick}}, every node/.append style={inner sep=0}]
		\node {
			\pgfimage[width=\x\textwidth]{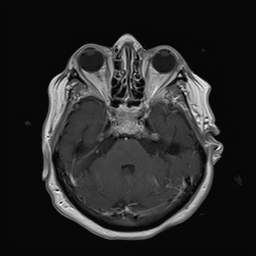}
		};
		\spy on (\locroiX, \locroiY) in node [left] at (\locmagX, \locmagY);     
	\end{tikzpicture}
	\caption{WDSR w/o GF}
\end{subfigure}	%
\hfil
\begin{subfigure}{0.19\textwidth}
	\begin{tikzpicture}[spy using outlines={rectangle,orange,magnification=2.5, 
		height=\y, width=\y, connect spies, /.append style={thick}}, every node/.append style={inner sep=0}]
		\node {
			\pgfimage[width= \x\textwidth]{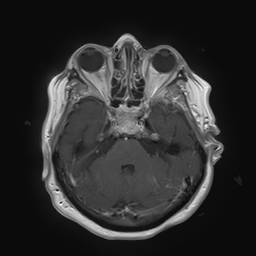}
		};
		\spy on (\locroiX, \locroiY) in node [left] at (\locmagX, \locmagY);     
	\end{tikzpicture}
	\caption{WDSR w/ GF}
\end{subfigure}	%
\hfil
\begin{subfigure}{0.19\textwidth}
	\begin{tikzpicture}[spy using outlines={rectangle,orange,magnification=2.5, 
		height=\y, width=\y, connect spies, /.append style={thick}}, every node/.append style={inner sep=0}]
		\node {
			\pgfimage[width= \x\textwidth]{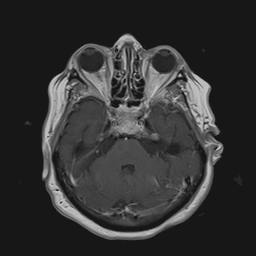}
		};
		\spy on (\locroiX, \locroiY) in node [left] at (\locmagX, \locmagY);     
	\end{tikzpicture}
	\caption{Ground truth}
\end{subfigure}	
\centering
%% PROJECTION IMAGES

\renewcommand\x{1}
\renewcommand\locroiX{0.2}
\renewcommand\locroiY{-0.8}
\renewcommand\locmagX{1.7}
\renewcommand\locmagY{0.7}
\hfil

\begin{subfigure}{0.19\textwidth}
	\begin{tikzpicture}[spy using outlines={rectangle,orange,magnification=2.5, 
		height=\y, width=\y, connect spies, /.append style={thick}}, every node/.append style={inner sep=0}]
		\node {
			\pgfimage[width= \x\textwidth]{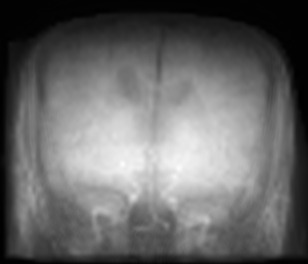}
		};
		\spy on (\locroiX, \locroiY) in node [left] at (\locmagX, \locmagY);     
	\end{tikzpicture}
	\caption{U-net w/o GF}
\end{subfigure}	%
\hfil
\begin{subfigure}{0.19\textwidth}
		\begin{tikzpicture}[spy using outlines={rectangle,orange,magnification=2.5, 
		height=\y, width=\y, connect spies, /.append style={thick}}, every node/.append style={inner sep=0}]
	\node {
		\pgfimage[width= \x\textwidth]{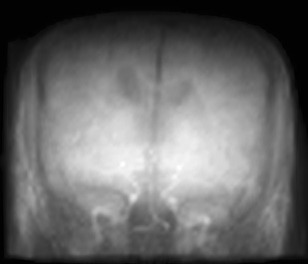}
	};
	\spy on (\locroiX, \locroiY) in node [left] at (\locmagX, \locmagY);     
	\end{tikzpicture}
	\caption{U-net w/ GF}
\end{subfigure}	%
\hfil
\begin{subfigure}{0.19\textwidth}
	\begin{tikzpicture}[spy using outlines={rectangle,orange,magnification=2.5, 
		height=\y, width=\y, connect spies, /.append style={thick}}, every node/.append style={inner sep=0}]
		\node {
			\pgfimage[width= \x\textwidth]{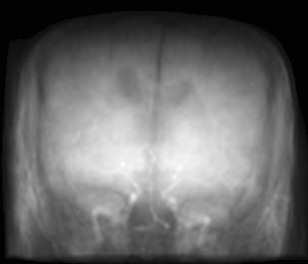}
		};
		\spy on (\locroiX, \locroiY) in node [left] at (\locmagX, \locmagY);     
	\end{tikzpicture}
	\caption{WDSR w/o GF}
\end{subfigure}	%
\hfil
\begin{subfigure}{0.19\textwidth}
	\begin{tikzpicture}[spy using outlines={rectangle,orange,magnification=2.5, 
		height=\y, width=\y, connect spies, /.append style={thick}}, every node/.append style={inner sep=0}]
		\node {
			\pgfimage[width= \x\textwidth]{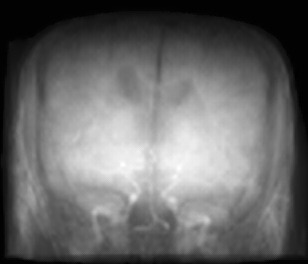}
		};
		\spy on (\locroiX, \locroiY) in node [left] at (\locmagX, \locmagY);     
	\end{tikzpicture}
	\caption{WDSR w/ GF}
\end{subfigure}	%
\hfil
\begin{subfigure}{0.19\textwidth}
	\begin{tikzpicture}[spy using outlines={rectangle,orange,magnification=2.5, 
		height=\y, width=\y, connect spies, /.append style={thick}}, every node/.append style={inner sep=0}]
		\node {
			\pgfimage[width= \x\textwidth]{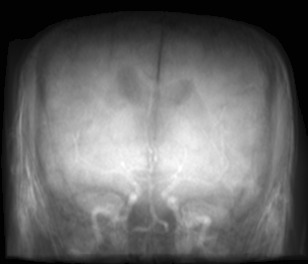}
		};
		\spy on (\locroiX, \locroiY) in node [left] at (\locmagX, \locmagY);     
	\end{tikzpicture}
	\caption{Ground truth}
\end{subfigure}	
	\caption{SR: Results for all network and guided filter configurations. Top row: tomographic T1 images. Bottom row: MR projection images.}
	\label{fig:preds_networks}
\end{figure*}

\begin{figure*}
	\input{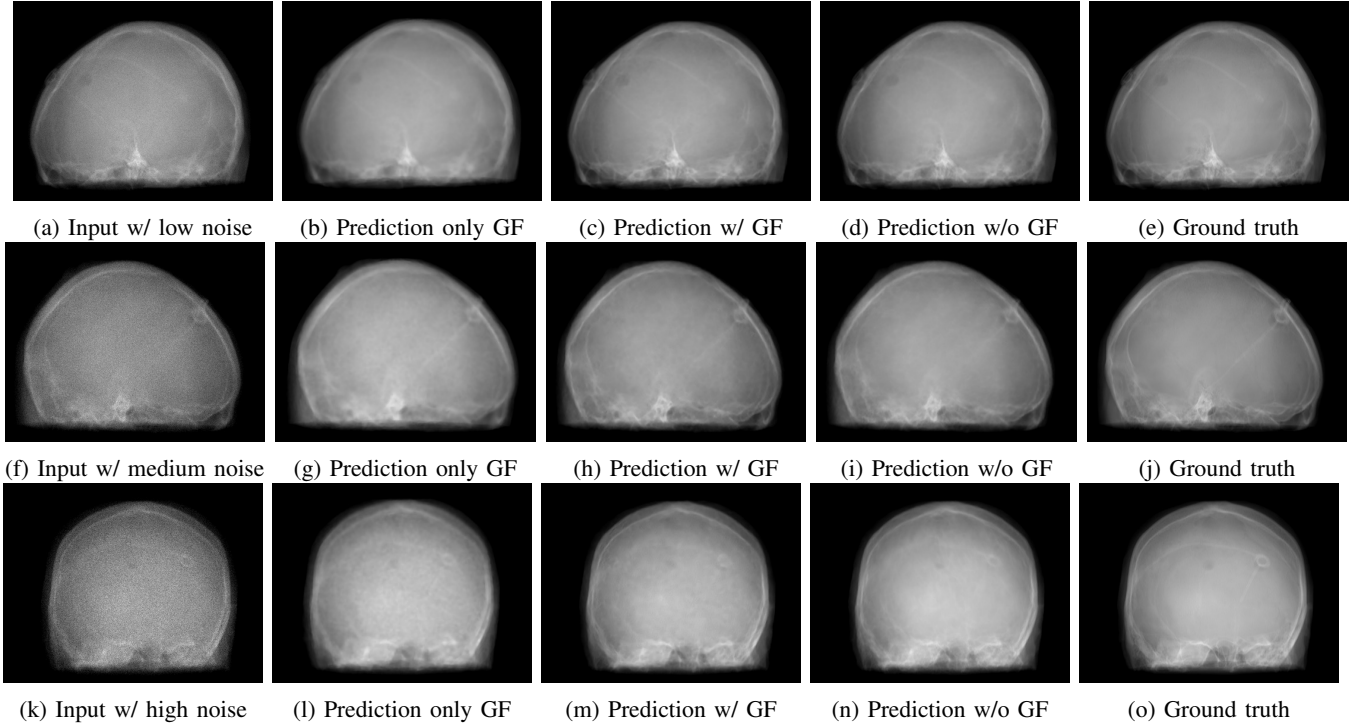}
	\caption{Denoising: Representative results for low (a)-(e), medium (f)-(j), and high (k)-(o) noise levels.}
	\label{fig:pipeline_io_denoising}
\end{figure*}

In Fig. \ref{fig:pipeline_io_sr} and \ref{fig:pipeline_io_denoising}, exemplary input, output and label images of the pipeline are presented for both tasks. Additional super-resolved images for both network architectures with and without the guided filter can be seen in Fig. \ref{fig:preds_networks}. The results show a consistently high quality over both, the tomographic and the projective domain, as well as both tasks. Corresponding quantitative evaluation can be found in Table \ref{tab:eval_patients} and \ref{tab:eval_patients_denoising}. For SR, the WDSR network, i.e., the designated super resolution network, performs consistently better for both datasets with and without the guided filter. Applied to the tomographic images, the approaches without the guided filter deliver slightly better quantitative results. For the projection images this difference diminishes and both approaches are on par. In the case of denoising, the approaches with the guided filter deliver a lower mean absolute error while the structural similarity is increased without it. Though, the measurable differences are only marginal. 
\red{The results generated by the plain guided filter without the learned guidance map are numerically worse than the approaches empowered by the guidance map generator for all tasks. This observation is most prevalent when observing the results of the tomographic T1 and T2 Flair images for SR, while for denoising the results are closer to the deep learning-based approaches.}

%Qualitative description SR
From a qualitative point of view, the results generated by the WDSR network exhibit slightly sharper visual appearance both with and without the guided filter for the case of super resolution. However, the observable deviations between all approaches are only minor. 
%Qualitative description Denoising
Larger deviations can be observed in the denoised images. With increasing the noise, the images denoised without the guided filter appear to be slightly smoother than those processed with it. \red{Here it becomes also apparent that the results generated only with the guided filter exhibit a blurred visual appearance and lack fine details.}

\begin{figure*}
	\centering
	\begin{subfigure}{0.32\textwidth}
		\includegraphics[width=1\textwidth]{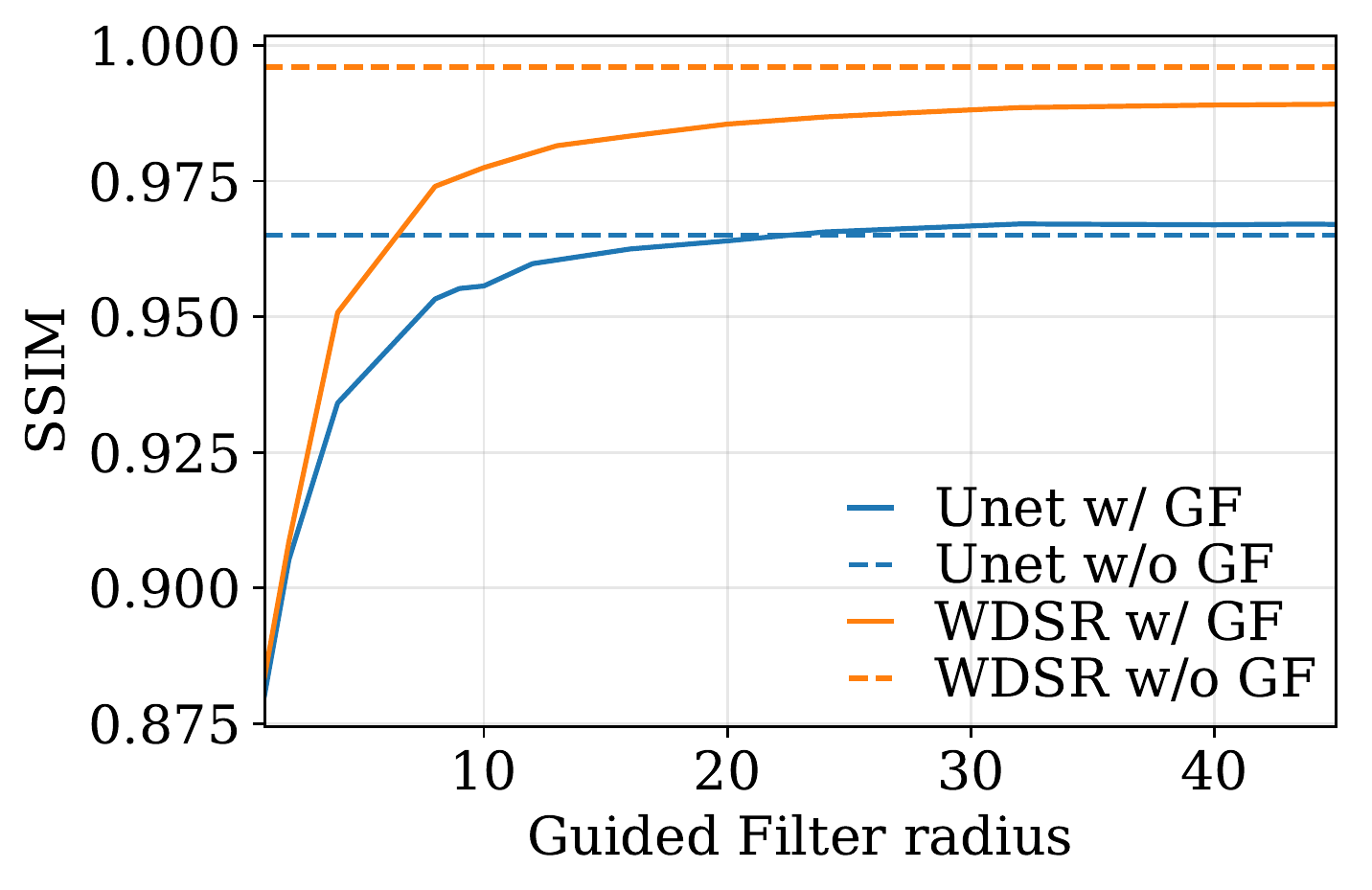}
		\caption{SR: tomographic data. }
		\label{fig:metrics_sr_radius_t1t2}
	\end{subfigure}
	\begin{subfigure}{0.32\textwidth}
		\includegraphics[width=1\textwidth]{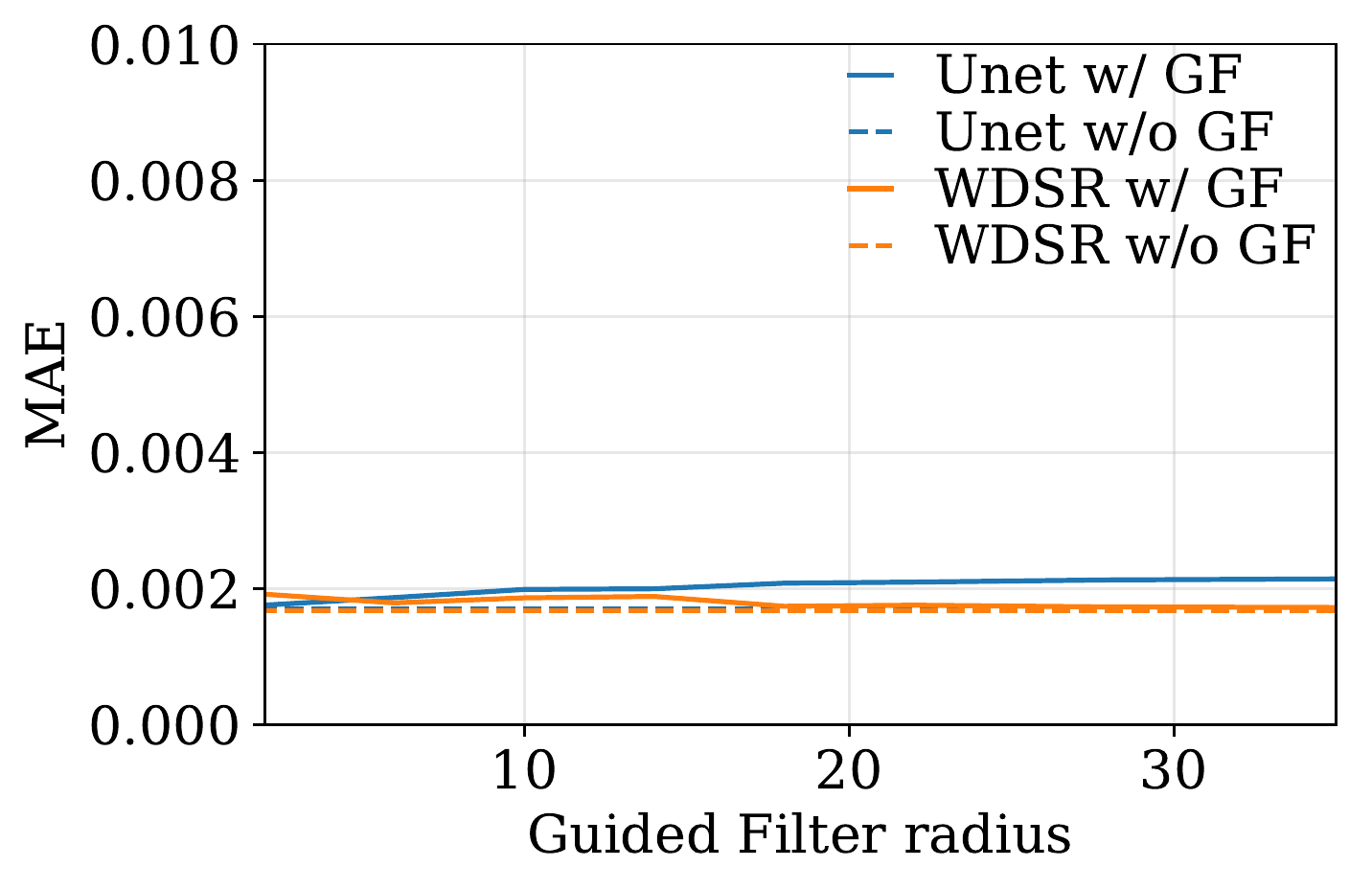}
		\caption{SR: projection data. }
		\label{fig:metrics_sr_radius_proj}
	\end{subfigure}
	\begin{subfigure}{0.32\textwidth}
		\includegraphics[width=1\textwidth]{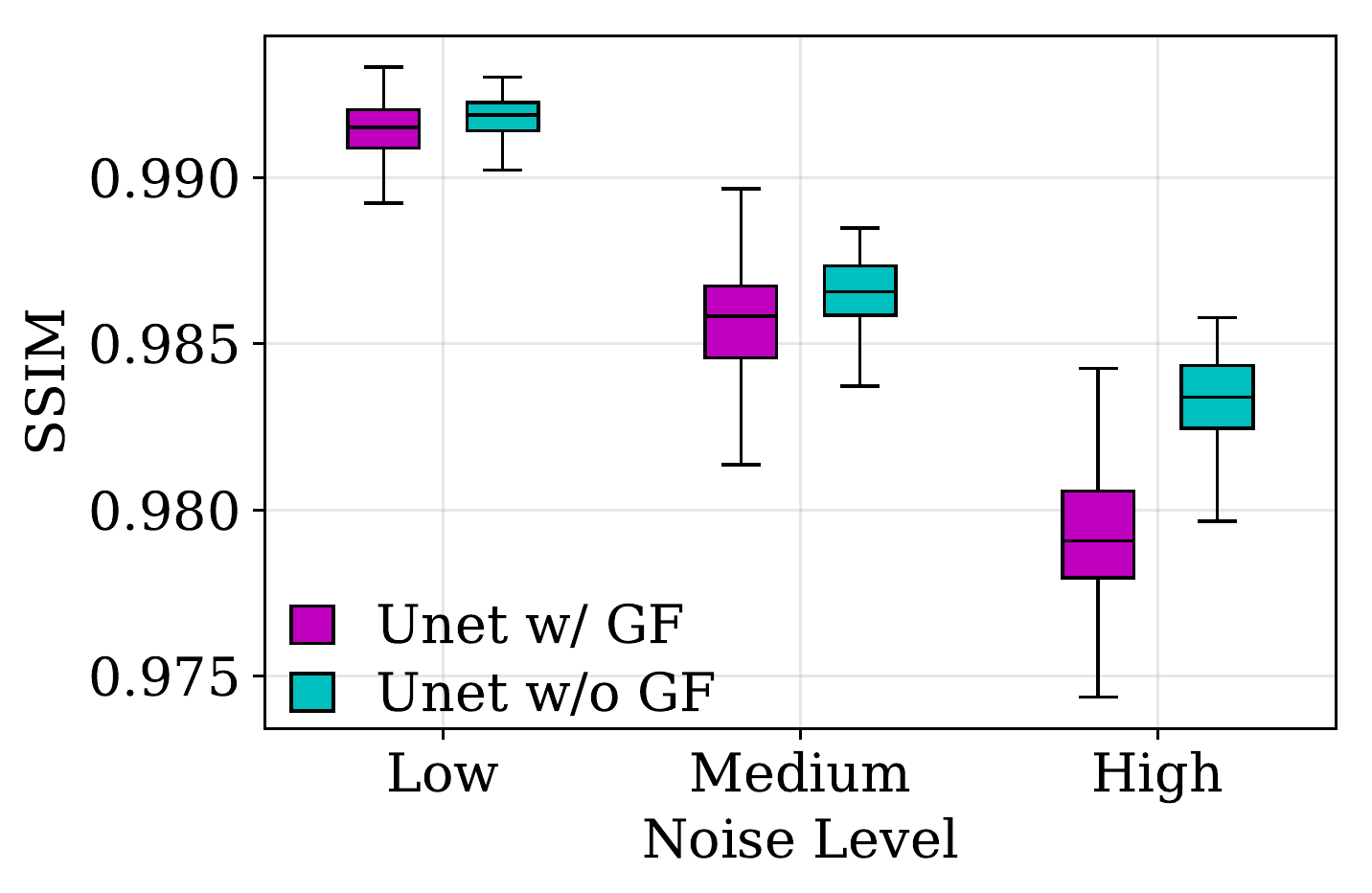}
		\caption{Denoising: projection data.}
		\label{fig:metrics_denoising}
	\end{subfigure}
	\caption{Quantitative metrics for super resolution w.r.t the radius of the guided filter (a)\,\&\,(b), and denoising of different noise levels (c).}
	\label{fig:metrics}
\end{figure*}

In Fig. \ref{fig:metrics_sr_radius_t1t2} and \ref{fig:metrics_sr_radius_proj}, the behavior of the evaluation metrics w.r.t. the chosen radius of the guided filter is presented for the super resolution case. For the tomographic T1 and T2 Flair image the error between the filtered output and the ground truth decreases with increasing radius. Applied on the higher resolution projection images, the quality of the output is almost constant w.r.t. to the radius. In accordance with \cite{Wu2018}, the epsilon parameter had only a negligible influence on the results in our experiments. 

\begin{figure}[]
	\centering
	\begin{subfigure}{0.45\textwidth}
		\includegraphics[width=1\textwidth]{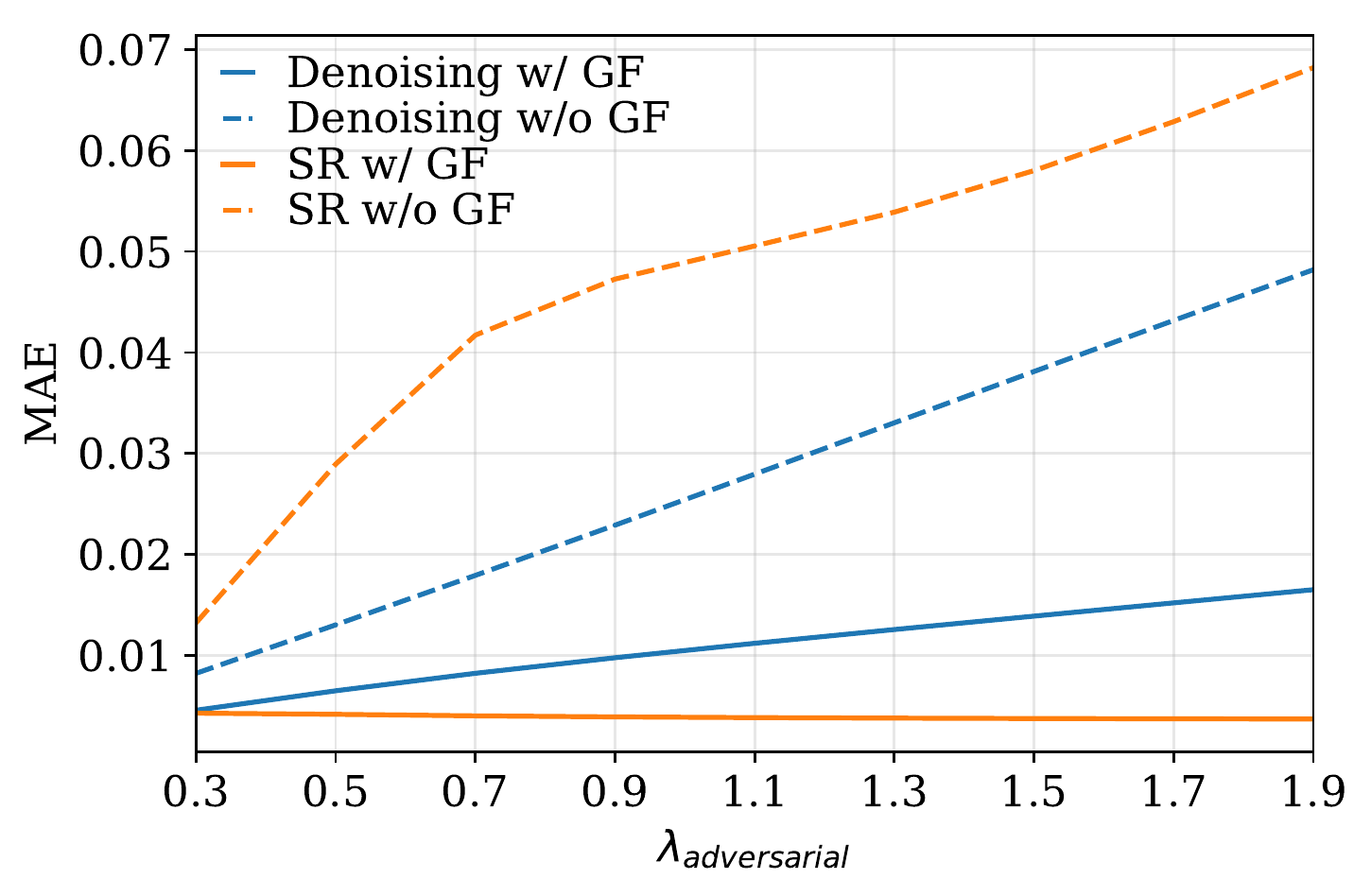}
		\caption{MAE for the adversarial attacks.}
		\label{fig:adv_lineplot}
	\end{subfigure}\\
	\begin{subfigure}{0.45\linewidth}
		\includegraphics[width=1\textwidth]{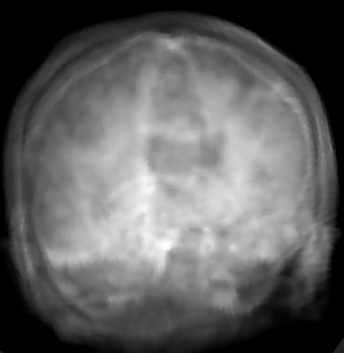}
		\caption{SR: attacked prediction w/ GF and  $\lambda_{\text{adversarial}}=1$.}
		\label{fig:adv_pred_wGF}
	\end{subfigure}
	\begin{subfigure}{0.45\linewidth}
		\includegraphics[width=1\textwidth]{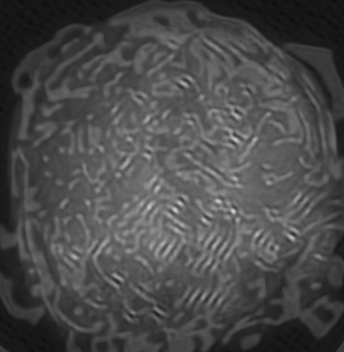}
		\caption{SR: attacked prediction w/o GF and  $\lambda_{\text{adversarial}}=1$.}
		\label{fig:adv_pred_woGF}
	\end{subfigure}
	\caption{Comprehensibility: Analysis of the vulnerability of the proposed pipeline to adversarial attacks as described in Sec. \ref{sec:exp_adversarial}. The pre-trained adversarial examples were normalized, rescaled by $\lambda_{\text{adversarial}}$, and subsequently added to the inputs $\vect{I}$ and $\vect{G}$ prior to inference.} %\red{[Bildbeschreibung vermutlich unverstaendlich. Formatierung evtl. anpassen.]}}
	\label{fig:adversarial_examples}
\end{figure}

\begin{figure}[t]
	\centering
	\includegraphics[width=1\linewidth]{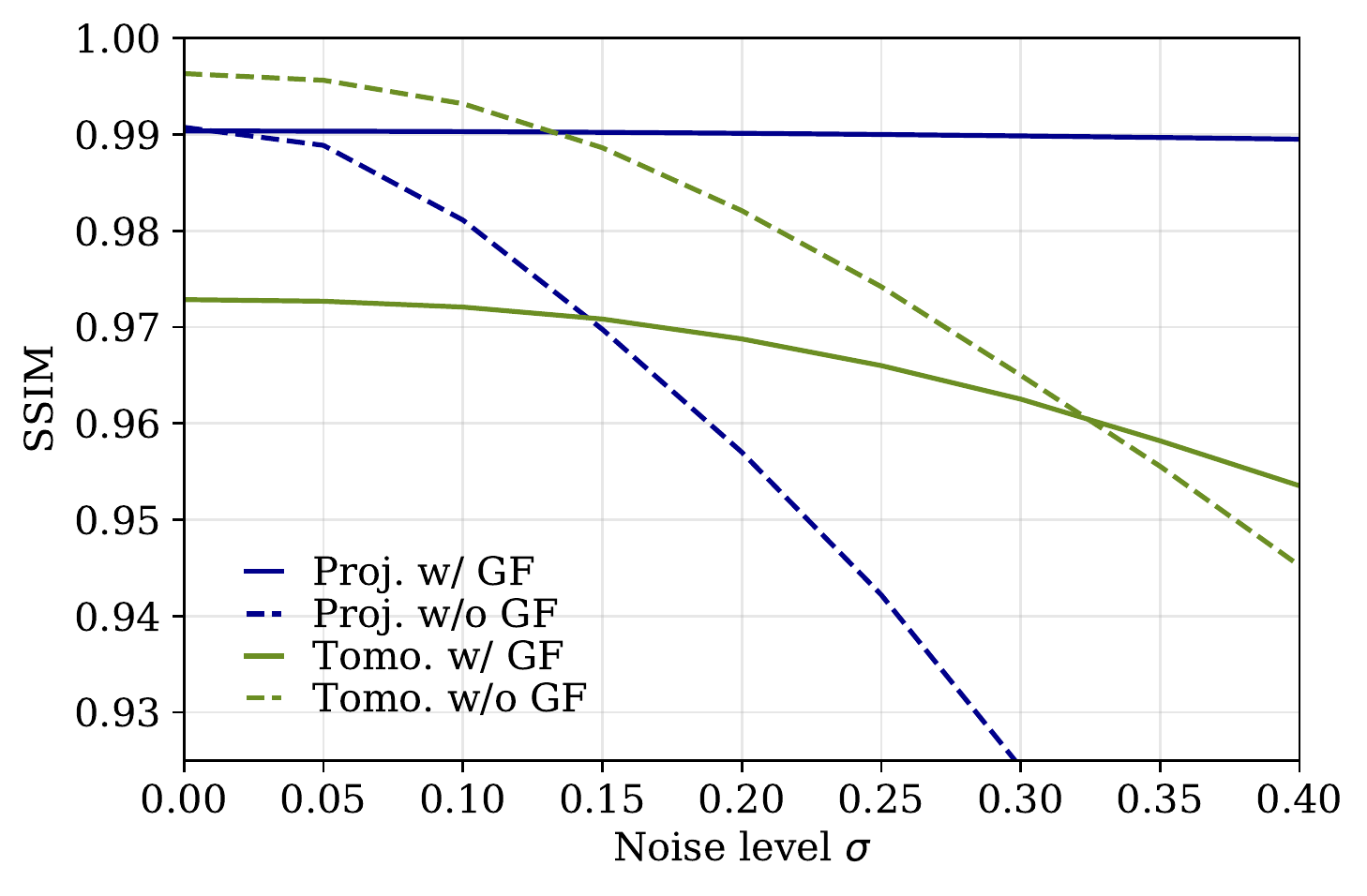}
	\caption{Comprehensibility: Investigation of the robustness against degradations. Plotted lines show the SSIM for increasing noise added to the guide image $\vect{G}$ in the SR case.}
	\label{fig:robustness}
\end{figure}

\begin{figure*}
	\centering
	\begin{subfigure}{0.45\textwidth}
		\includegraphics[width=1\textwidth]{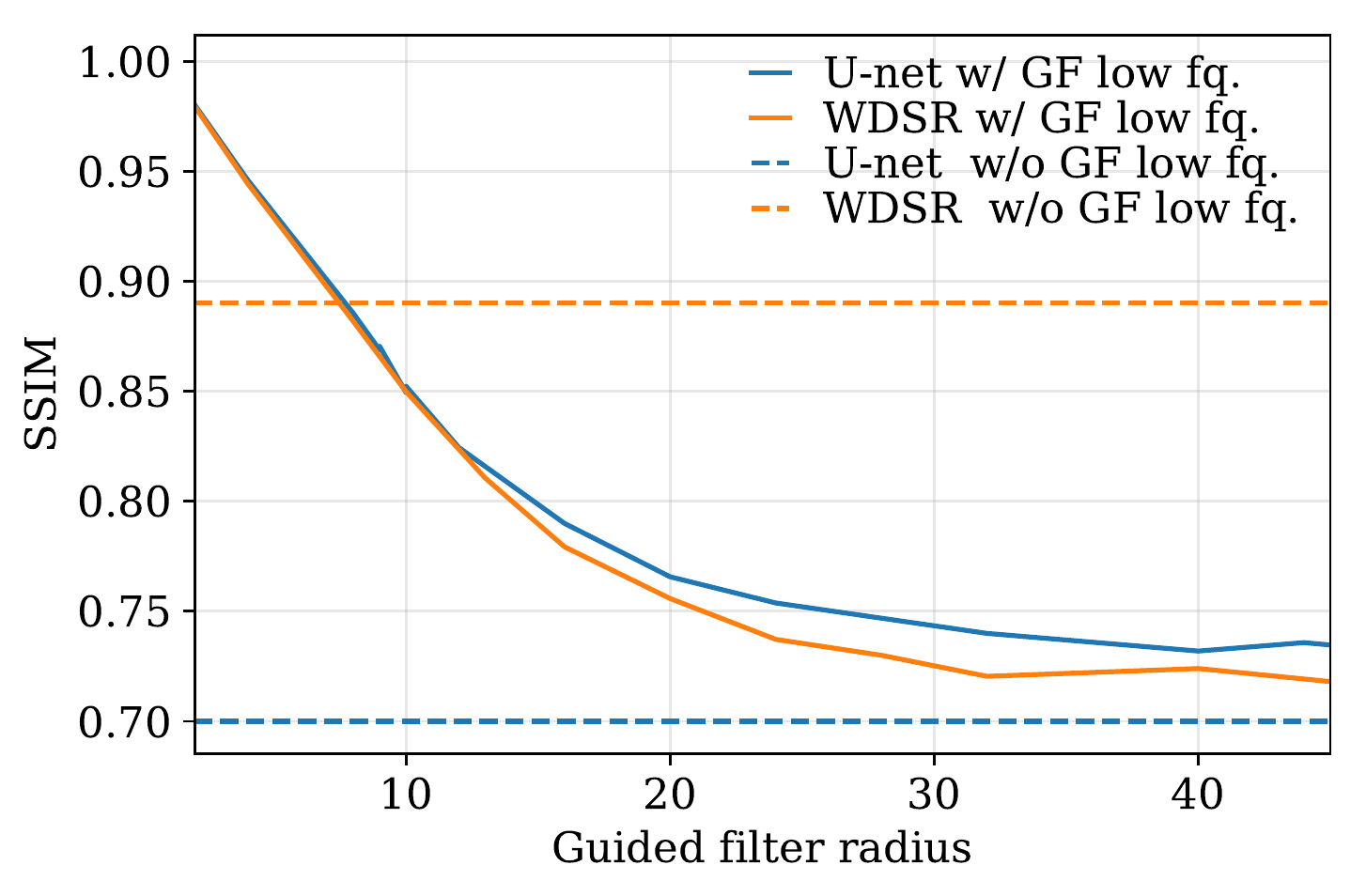}
		\caption{SR: low-frequency analysis.}
		\label{fig:low_fq_analysis_sr}
	\end{subfigure}
	\begin{subfigure}{0.45\textwidth}
		\includegraphics[width=1\textwidth]{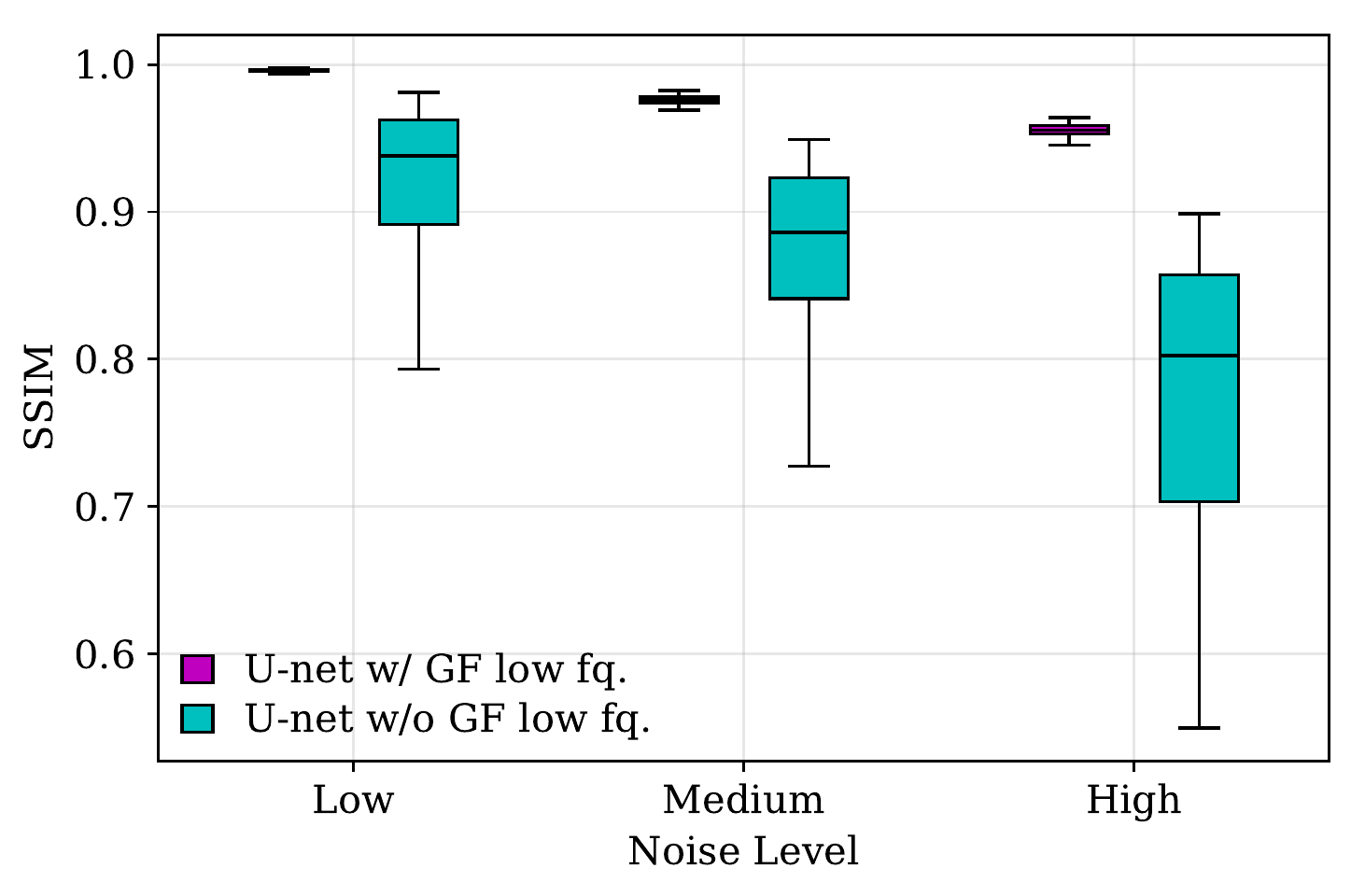}
		\caption{Denoising: low-frequency analysis.}
		\label{fig:low_fq_analysis_denoising}
	\end{subfigure}
	\caption{Comprehensibility: Analysis of the change in the images' content as described in Sec. \ref{sec:exp_content_pres}. Left: low frequency components for SR w.r.t. the radius of the guided filter. Right: low frequency components for different denoising levels.}
	\label{fig:low_fq_analysis}
\end{figure*}

\textit{Comprehensibility}:
The experiments towards the comprehensibility of the results substantiate our claim that the guided filter is able to decouple the neural network from the output which is an important step towards secure medical image processing. In Fig. \ref{fig:low_fq_analysis} it can be seen that the images' content in the form of low frequency components is better preserved when using the guided filter than without it. Though, this is subject to the chosen radius of the guided filter. With increasing radius the structural similarity of the low frequency sub-bands decreases which indicates larger changes in the images' content. Accordingly, the results used for the presented evaluation were created with a small guided filter radius of 8 for the tomographic (SR) and 2 for the projection images (SR \& Denoising). 
Fig. \ref{fig:robustness} shows the behavior of the approaches for distorted input data. It becomes apparent that the pipeline with the guided filter is much more robust against disturbance of the high-resolution guide images where as without it the degradations are transferred to the output image. Similarly, the presented error plot in Fig. \ref{fig:adv_lineplot} shows that the guided filter is able to limit the effectiveness of an adversarial attack on the output images. This effect increases with increasing severity of the attack.

Additional results for both tasks and all network configurations are presented in Appendix E.

\section{Discussion}
\label{sec:Discussion}
\begin{figure*}
	\centering
	\begin{subfigure}[t]{0.245\textwidth}
		\includegraphics[width=1\textwidth]{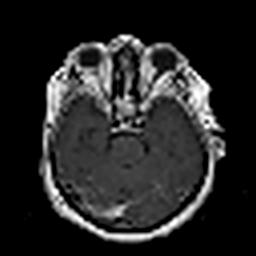}
		\caption{Upsampled Input $\vect{I}_{\text{up}}$}
		\label{fig:shutout_input}
	\end{subfigure}
	\begin{subfigure}[t]{0.245\textwidth}
		\includegraphics[width=1\textwidth]{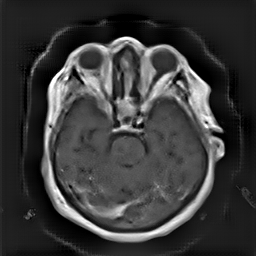}
		\caption{Guidance map $\vect{M}$\,w/o\,guide $\vect{G}$}
		\label{fig:shutout_wo_guide}
	\end{subfigure}
	\begin{subfigure}[t]{0.245\textwidth}
		\includegraphics[width=1\textwidth]{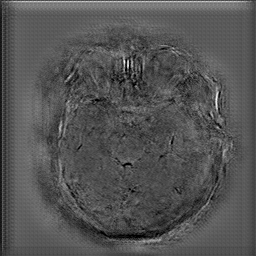}
		\caption{Guidance map $\vect{M}$\,w/o\,\mbox{input} $\vect{I}$}
		\label{fig:shutout_wo_input}
	\end{subfigure}
	\begin{subfigure}[t]{0.245\textwidth}
		\includegraphics[width=1\textwidth]{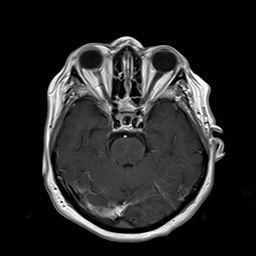}
		\caption{Ground truth}
		\label{fig:shutout_label}
	\end{subfigure}
	\caption{Assessment of the composition of the guidance map for the super resolution task. The figure shows the resulting guidance maps generated by the trained model without the second high-resolution modality (b) or without the low-resolution input (c), respectively. In (b) it can be observed that the image content is mainly drawn from the low-resolution input while (c) shows that the second guide modality is used almost exclusively to generate the high-frequency details like edges. Note that we assume the square pattern in (c) to originate from the periodic pixel shuffling employed by the WDSR network. }
	\label{fig:shutout}
\end{figure*}
The conducted evaluation substantiates that the combination of deep learning and known operators, in this case the guided image filter, is capable to produce results that are on par with state-of-the-art approaches. More importantly, the proposed approach makes it possible to process the input images without altering their general content. 
As a result, even in areas with sensitive data, the power of deep learning can be used without having to accept unpredictable changes to the information contained. 

We performed three experiments to substantiate the increased comprehensibility of the results created with the guided filter. 
\subsubsection{Discussion: Content Preservation}
\label{sec:disc_content_preservation}
First, we investigated the change of the input image's underlying content w.r.t. the predicted outputs (see Fig. \ref{fig:low_fq_analysis_sr} and \ref{fig:low_fq_analysis_denoising}). For small radii of the guided filter, the signal fidelity between these images is considerably higher with the guided filter than without it in the super resolution case. With increasing radii the error decreases, however, simultaneously the deviation of the output from the input signal increases in the low frequencies. This indicates, as described in Section \ref{sec:exp_content_pres}, an increasing change of the image content with increasing influence of the guidance map. This holds for both cases, SR and denoising. In Fig. \ref{fig:low_fq_analysis_denoising} it can be seen that the outputs generated with the guided filter not only exhibit higher structural similarity in average but also drastically lower variation than those processed only by the neural network. With increasing noise, i.e., a higher proportion of hallucinated information by the network in the output images, this effect becomes even stronger.

\subsubsection{Discussion: Insusceptibility to Disturbances}
\label{sec:disc_disturbances}
Second, the behavior of the pipeline for distorted guide images was analyzed. The results presented in Fig. \ref{fig:robustness} show that for increasing noise the guidance map degrades steadily. The guided filter is able to limit the influence of the guidance map in this case whereas the plain network has no means to preserve the input signal. Training the networks to cope with the noisy images of this simple experiment is possible. However, in real world applications, the variety and severity of possible degradations of the input signals is enormous. While covering all possibilities in the training process is infeasible, the guided filter is able to inherently provide protection against these flaws. 

\subsubsection{Discussion: Robustness against Adversarial Attacks}
\label{sec:disc_adversarial}
Third, we could show that the guided filter also provides an efficient defense against adversarial attacks. The attacked prediction created with the guided filter (Fig. \ref{fig:adv_pred_wGF}) exhibits a blurred but still natural appearance of the output image. In contrast, without the guided filter, the attacked prediction (Fig. \ref{fig:adv_pred_woGF}) is heavily deteriorated. Furthermore, in Appendix D, training statistics from the creation of the adversarial attacks are included. These substantiate that it is clearly harder to train suitable attacks in the first place when using the guided filter. 
\newline
The conducted experiments towards the comprehensibility strongly indicate that decoupling the network from the input signal using the guided filter yields more robust and trustworthy outputs compared to deep learning techniques alone. In all considered cases the guided filter could limit the manipulation of the input image's content and additionally provide a defensive mechanism against degradation. In order to follow up on to these findings, we hope to develop further tests regarding the comprehensibility of the results in the future. We would also like to encourage other researchers to evaluate their own methods for comprehensibility.
\newline

In general, the decrease in image quality metrics w.r.t. to smaller radii is much stronger for the tomographic T1 and T2 Flair images than for the projection images. We assume the reason for this to be the difference in resolution of both datasets. 
The projection images resemble the ground truth more closely from the outset when compared to the tomographic images as seen in Fig. \ref{fig:pipeline_io_t1t2_input} and \ref{fig:pipeline_io_proj_input}, respectively.
Consequently, more information has to be generated by the networks. As the guided filter is sensitive towards correlation between the input and guidance map the constraints on the applicable changes are more severe. 

In Fig. \ref{fig:shutout}, the composition of the guidance map from the multiple input images to the generator network is assessed for the super resolution task. As intended, the main part of the predicted image's content is drawn from the input image, while only high-resolution details and edges are extracted from the guide image. This substantiates that valuable information is present in the complementary guide images.
Currently only pre-registered input and guide images are considered. In future work it is therefore necessary to investigate the influence of registration errors on the outcome of the pipeline, especially for parts of the body that can be subject to larger soft tissue deformation, e.g., the abdomen. 

Differences between the methods with and without the guided filter are measurable but hardly visible. An important aspect is that the better performing network in combination with the guided filter delivers consistently better results than the worse network without the guided filter (see Tab. \ref{tab:eval_patients} and Fig. \ref{fig:preds_networks}). I.e., improvements on the network side are preserved throughout the proposed guided filtering pipeline. Consequently, future advances in deep learning methods can easily contribute to this approach. The modular structure of the proposed guided filtering pipeline allows to easily exchange the chosen guidance map generator without further complications.

\section{Conclusion}
\label{sec:conclusion}
We proposed a pipeline for multi-modal image processing that combines the advantages of deep learning and the guided image filter and provided detailed analysis of the comprehensibility of the results.
The optimal mapping of multi-modal inputs to form the guidance map for the guided filter is learned directly from data by a neural network. The output itself, however, is only processed by the locally linear operations of the guided image filter. Through this combination, the incomprehensible transformation of the neural network is decoupled from the output.
The analysis based on two showcase tasks, image super resolution and denoising, showed that the generated results with the guided filter are on par with the state-of-the-art approaches quantitatively as well as qualitatively. Opposed to conventional methods of deep learning, this can be done with considerably less manipulation of the underlying image's content information. Furthermore, the proposed pipeline is more robust against degradations of the input as well as adversarial attacks. 
In conclusion, the proposed approach allows for the comprehensible application of deep learning-based image processing in areas with low tolerance for errors or imponderables such as medical imaging.

\bibliographystyle{IEEEtran}

%%%%%%%%%%%%%%%%%%%%%%%%%%%%%%%%%%%%%%%%%%%%%%%%
%%% Appendix%%%%%%%%%%%%%%
%%%%%%%%%%%%%%%%%%%%%%%%%%%%%%%%%%%%%%%%
\newpage
\appendices
\onecolumn
\section{Super Resolution \& Denoising: U-Net}
\begin{figure*}[h]
	\centering
	\includegraphics[width=1\textwidth]{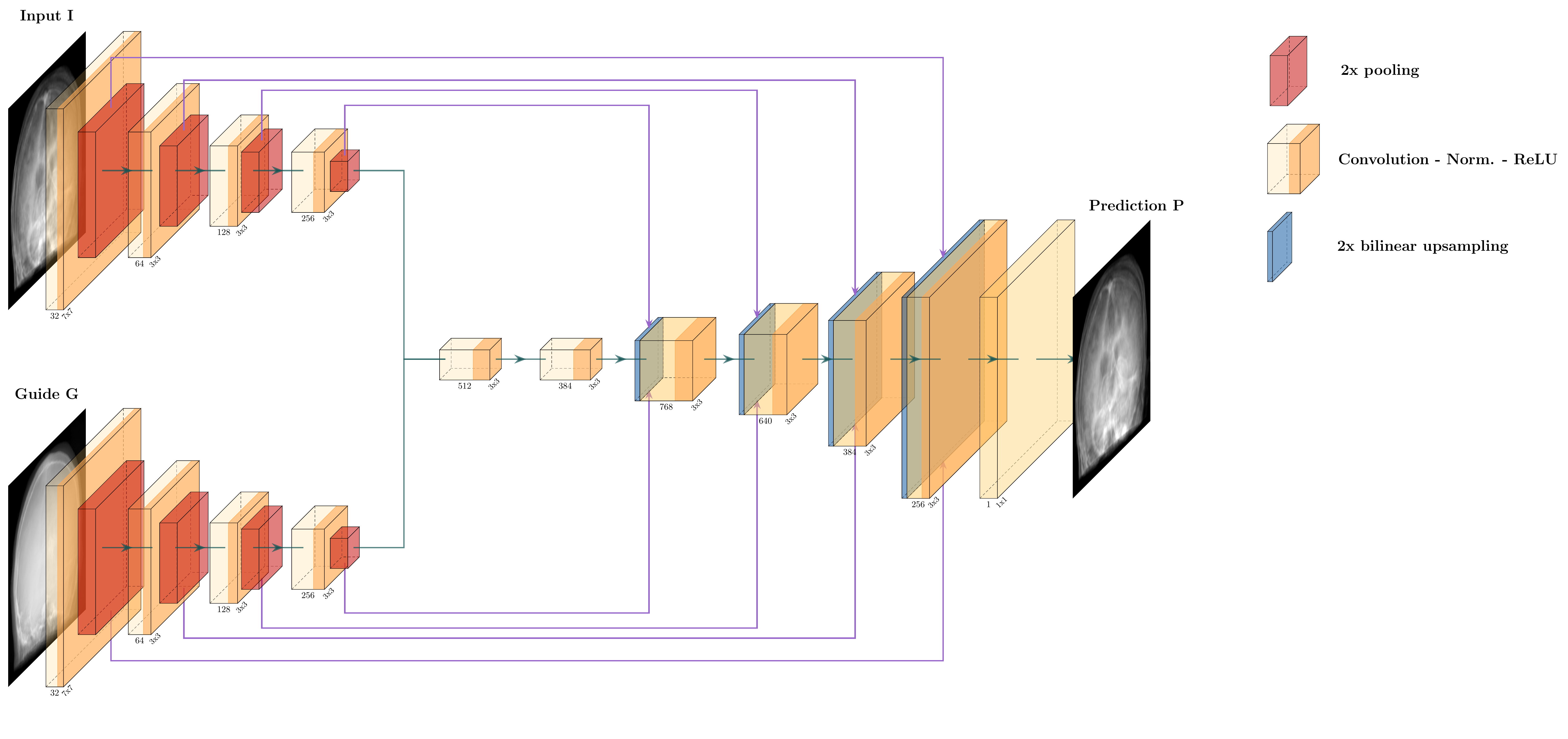}
	\caption{Graphical representation of the employed U-Net network.}
	\label{fig:unet}
\end{figure*}

Fig. \ref{fig:unet} provides a representation of the U-Net that is used for super resolution and denoising. The general architecture is designed according to \cite{Ronneberger2015}, implementation by \footnote{\url{https://github.com/milesial/Pytorch-UNet}}. Our modification is confined to adding the encoding branch for the guide image $\vect{G}$ (bottom) which is equivalent to the input image's encoding branch and replacing batch normalization by instance normalization \cite{Ulyanov2016}. The batch size and learning rate are set to 1 and $1e^{-5}$, respectively (see main paper, Sec. III).

\section{Super Resolution: WDSR}
\begin{figure*}[]
	\centering
	\includegraphics[width=1\textwidth]{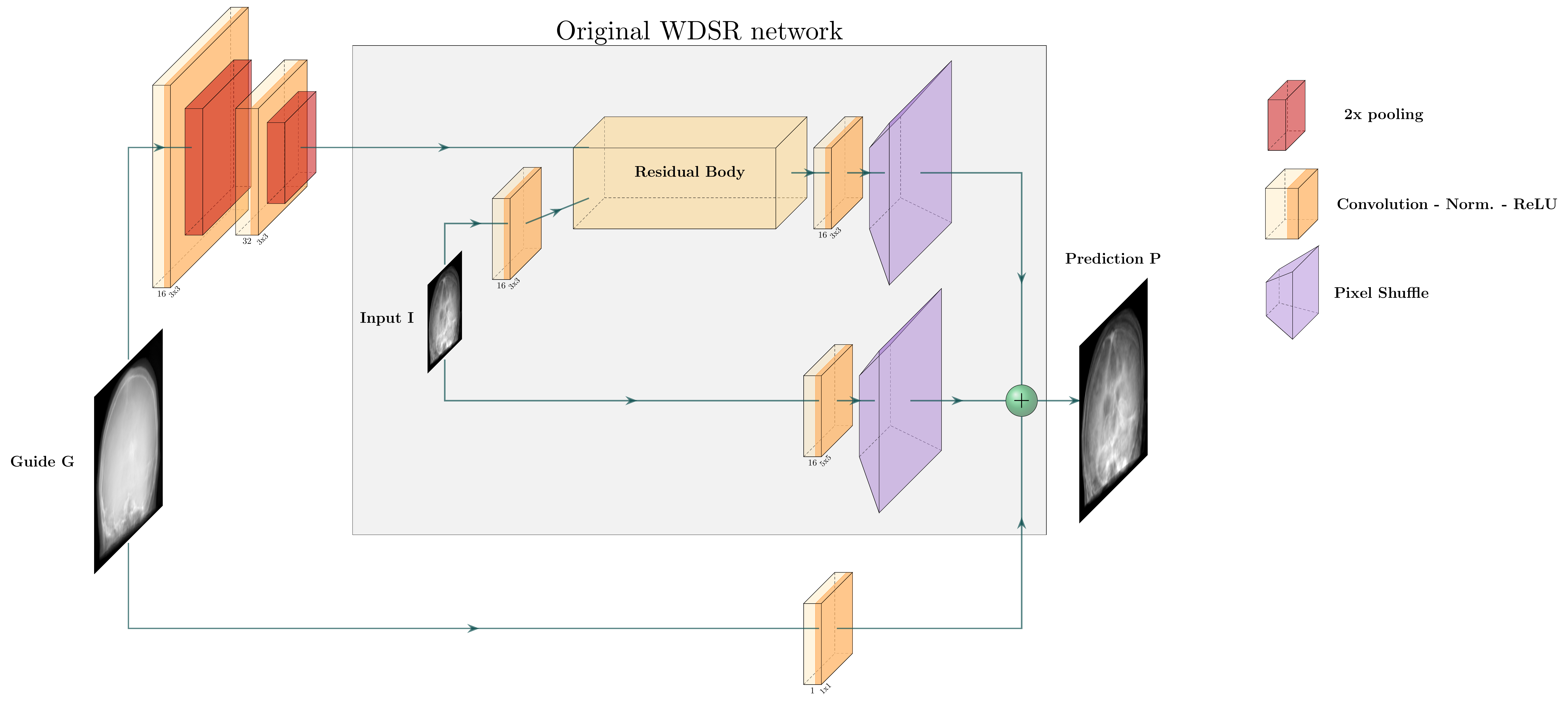}
	\caption{Graphical representation of the modified WDSR network. Visualization based on \cite{Yu2018}.}
	\label{fig:wdsr}
\end{figure*}

Fig. \ref{fig:wdsr} shows the WDSR network that is used for super resolution. The original WDSR network \cite{Yu2018, Fan2018} \footnote{\url{https://github.com/JiahuiYu/wdsr_ntire2018}} is highlighted by the gray box. 
Solely the operations targeting color images were removed. The batch size and learning rate were set to 1 and $1e^{-5}$ as described in Sec. III of the main paper. All other parameters were kept unchanged (configuration: wdsr\_a) except in the case of projection images where the number of residual blocks in the residual body was reduced from 32 to 24 to account for memory limitations. To accept two inputs modalities, a second residual connection is added (bottom). Furthermore, given the higher initial resolution of the guide image, the original network is preceded by two convolutional encoding blocks as used in the U-net to match both input images' spatial resolution.

\section{Comprehensibility: Content Preservation}
\label{sec:appendix_wavelets}
\begin{figure*}[]
	\centering
	\includegraphics[width=0.7\textwidth]{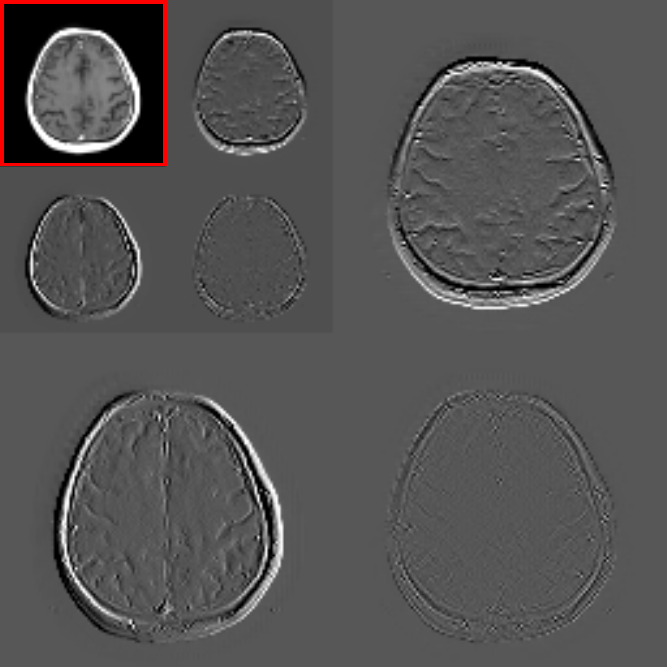}
	\caption{An example of the wavelet decomposition resulting from the experiment towards comprehensibility. The top left image (marked in red) contains the low-frequency components of the two-fold wavelet decomposition of the predicted super-resolved image. Note that the images' windowing is selected individually to ensure appropriate visibility.}
	\label{fig:wavelet}
\end{figure*}

In Fig. \ref{fig:wavelet}, a two-fold wavelet decomposition (Daubechies-4 \cite{Daubechies1992}) of the predicted super-resolved is presented. It can be observed that most of the image's content information is contained in this image. Ideally, these low-frequency components are as close as possible to the low-resolution input image as we do not intend to change this content information. The remaining images show the high-resolution components which are subject to change in the super-resolution process. 

\clearpage
\section{Comprehensibility: Adversarial Attack}
\label{sec:appendix_adversarial}
\begin{figure}[h]
	\centering
	\includegraphics[width=0.5\textwidth]{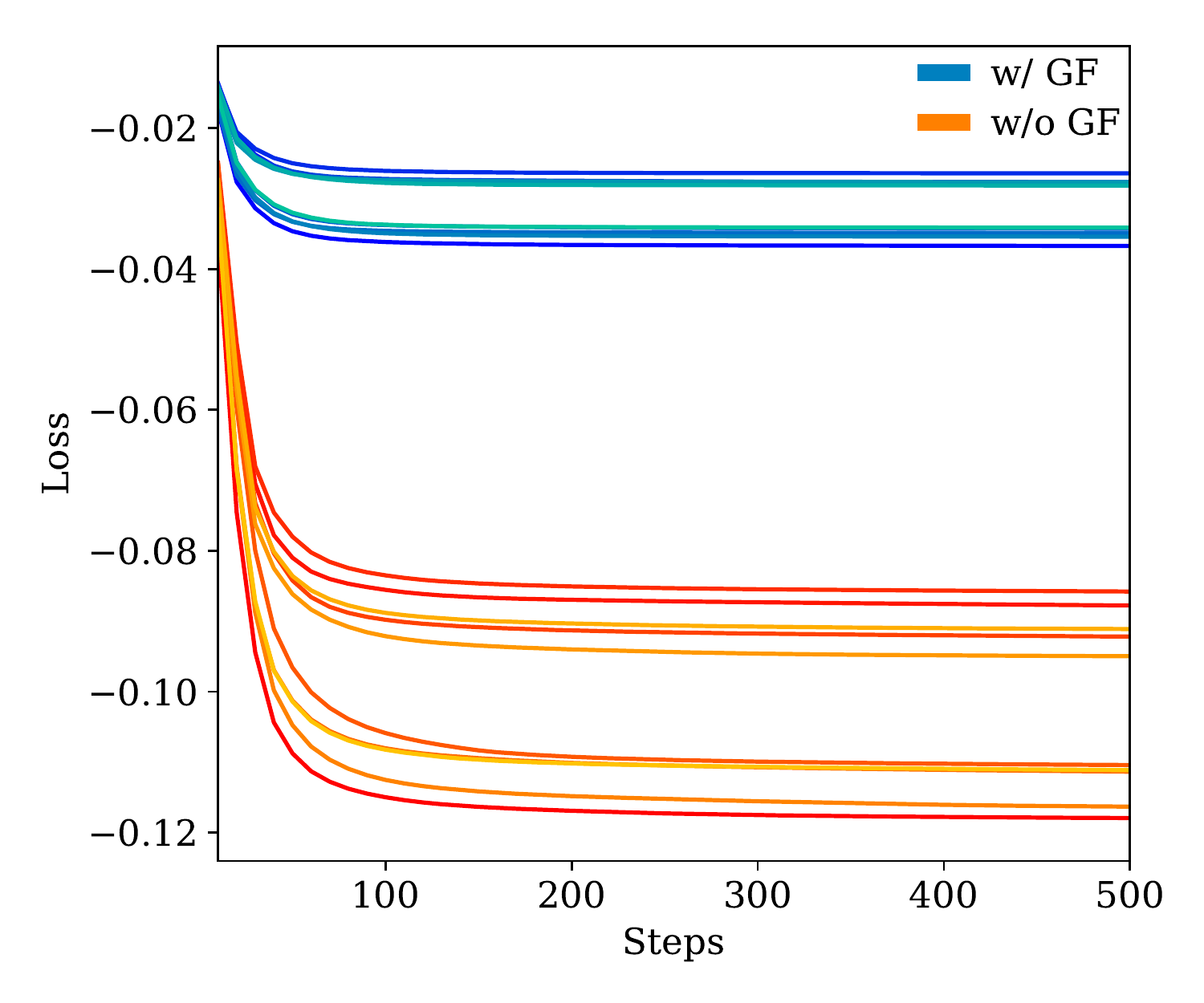}
	\caption{Loss statistics of the training of adversarial examples. Both setups, with and without the guided filter, were completely equivalent.}
	\label{fig:adv_training_plot}
\end{figure}

To perform an adversarial attack on the networks suitable adversarial examples are required. To this end, we need to find two adversarial examples $\vect{E}_I$ and $\vect{E}_G$ that are added to the input $\vect{I}$ and guide $\vect{G}$ images. These added residuals should exhibit two properties. First, when added they should cause a large deviation of the predicted output from the real ground truth, here denoted as $\vect{L}$. Second, simultaneously, the adversarial examples should apply only minimal changes to the original inputs. This is enforced by penalizing the magnitude of the Euclidean norm of the adversarial examples. Starting with a pre-trained guidance map generator employing the transformation $\phi$ that maps the input images to the desired output, the predicted image $\vect{P}$ is given $\vect{P} = f_{\text{GF}}(\phi(\vect{I} + \vect{E}_I, \vect{G} + \vect{E}_G), \vect{I})$ when using the guided filter, or directly by  $\vect{P} = \phi(\vect{I} + \vect{E}_I, \vect{G} + \vect{E}_G)$ otherwise. The resulting optimization problem is then defined by
\begin{equation}
\min |\vect{L} - \vect{P}| + \lambda \left(\lVert \vect{E}_I \rVert_2 + \lVert \vect{E}_G \rVert_2 \right) \;.
\end{equation}
We solve this optimization problem in the same framework as the initial training of the guidance map generator using the ADAM optimizer~\cite{Kingma2014} with a learning rate of $1\text{e}^{-2}$ and learning rate decay down to $1\text{e}^{-5}$.
In Fig. \ref{fig:adv_training_plot} loss statistics for the training of the adversarial examples are presented. During training, the setups for both approaches, with and without the guided filter, were completely equivalent. It becomes apparent that the guided filter not only provides an effective defense against adversarial examples as seen in the results section of the main paper, but also aggravates training of suitable adversarial examples in the first place. 

\clearpage
\newpage
\section{Additional Super-Resolved and Denoised Images}
\label{sec:appendix_results_sr}

\begin{figure*}[h]
	\input{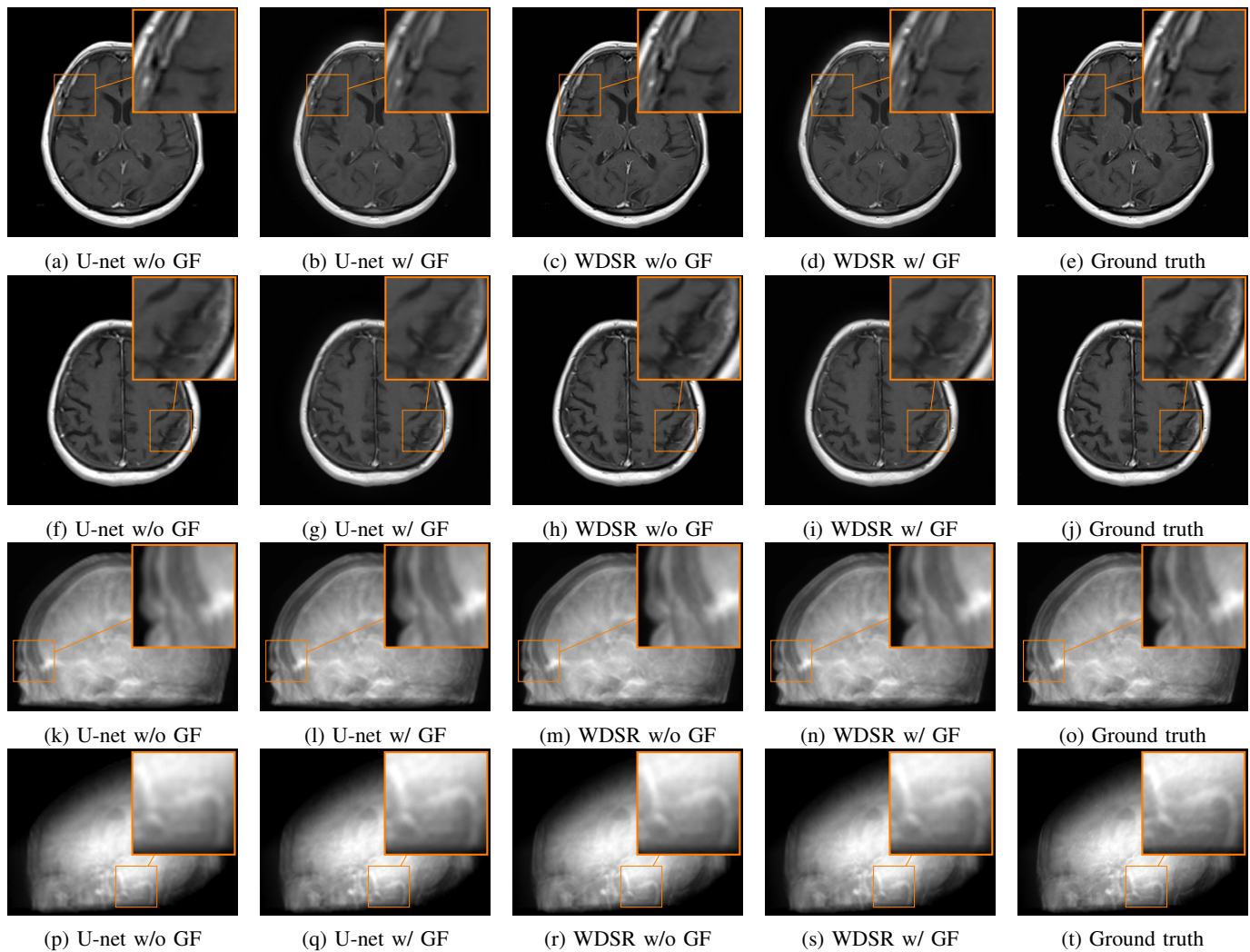}
	\caption{SR: Results for all network and guided filter configurations. First and second row: tomographic T1 images. Third and fourth row: MR projection images.}
	\label{fig:preds_networks_sr}
\end{figure*}
\clearpage
\begin{figure*}[t!]
	\input{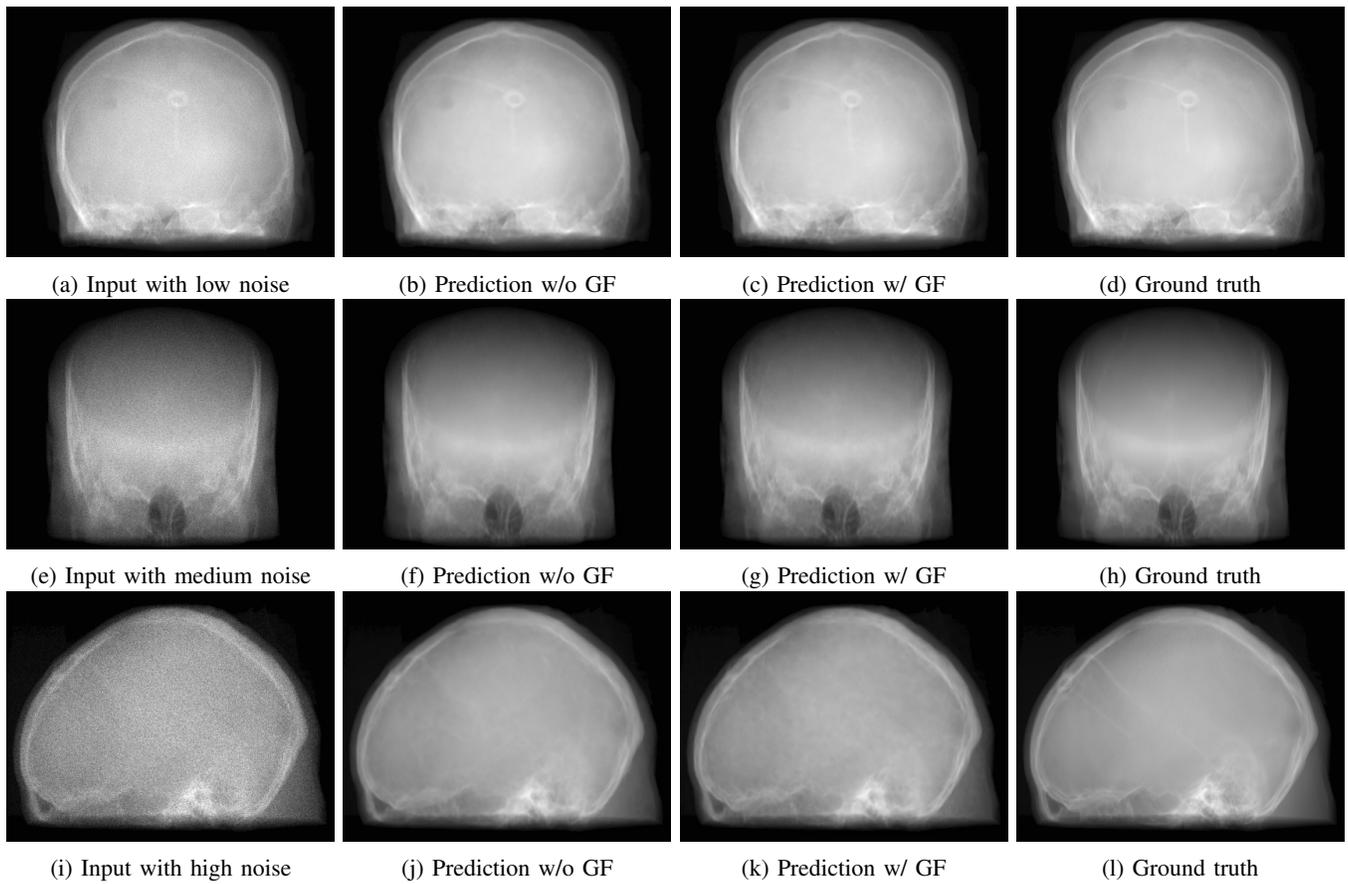}
	\caption{Denoising: Additional results for all noise levels with and without the guided filter.}
	\label{fig:preds_networks_denoising}
\end{figure*}

\end{document}